\documentclass[]{revtex4-2}

\usepackage{amsfonts}
\usepackage{amsmath}
\usepackage{graphicx}
\usepackage{bbm}
\usepackage{float}
\usepackage{fullpage}
\usepackage{color}

\DeclareMathOperator*{\argmin}{arg\,min}

\usepackage{lineno,hyperref}
\modulolinenumbers[5]

\begin{document}

% \begin{frontmatter}

\title{Predicting rare events using neural networks and short-trajectory data}
%\title{Elsevier \LaTeX\ template\tnoteref{mytitlenote}}

%\author{John Strahan\fnref{myfootnote}}
%\address{Department of Chemistry, University of Chicago, Chicago, IL 60637}

% %% or include affiliations in footnotes:
% \author[mymainaddress]{John Strahan}
% %\ead{support@elsevier.com}

% \author[eaps]{Justin Finkel}
% %\ead{support@elsevier.com}

% \author[mymainaddress,ccam]{Aaron R.\ Dinner}
% \cortext[mycorrespondingauthor]{Corresponding author}
% \ead{dinner@uchicago.edu}

% \author[courant]{Jonathan Weare\corref{mycorrespondingauthor}}
% \cortext[mycorrespondingauthor]{Corresponding author}
% \ead{weare@nyu.edu}

% \address[mymainaddress]{Department of Chemistry and James Franck Institute, the University of Chicago, Chicago, IL 60637}
% \address[ccam]{Committee on Computational and Applied Mathematics, the University of Chicago, Chicago, IL 60637}
% \address[courant]{Courant Institute of Mathematical Sciences, New York University, New York, New York 10012}

% \address[eaps]{Department of Earth, Atmospheric, and Planetary Sciences, Massachusetts Institute of Technology, Cambridge, MA 02139}

\author{John Strahan}
\affiliation{Department of Chemistry and James Franck Institute, University of Chicago, Chicago IL 60637, USA}
\author{Justin Finkel}
\affiliation{Committee on Computational and Applied Mathematics, the University of Chicago, Chicago, IL 60637 USA}
\author{Aaron R. Dinner}
\email{dinner@uchicago.edu}
\affiliation{Department of Chemistry, James Franck Institute, and Committee on Computational and Applied Mathematics, University of Chicago, Chicago IL 60637, USA}
\author{Jonathan Weare}
\email{weare@nyu.edu}
\affiliation{Courant Institute of Mathematical Sciences,\\ New York University, New York 10012, USA}

\begin{abstract}
Estimating the likelihood, timing, and nature of events is a major goal of modeling stochastic dynamical systems.  When the event is rare in comparison with the timescales of simulation and/or measurement needed to resolve the elemental dynamics, accurate prediction from direct observations becomes challenging.
%A common task in the analysis of stochastic dynamical systems is forecasting.  
%Examples of this include computing the folding probability of large molecular systems, or calculating the average time until the emergence of a rare, severe weather event in an atmospheric model.  
In such cases a more effective approach is to cast statistics of interest as solutions to Feynman-Kac equations (partial differential equations).
%A common approach to solving such forecasting problems is to cast the desired statistics as solutions to Feynman-Kac equations.  
Here, we develop an approach to solve Feynman-Kac equations by training neural networks on short-trajectory data.  
{ Our approach is based on a Markov approximation but 
otherwise avoids assumptions about the underlying model and dynamics.}
% , unlike previous methods.}  
This makes it applicable to treating complex computational models and observational data.   We illustrate the advantages of our method using a low-dimensional model that facilitates visualization, and this analysis motivates
%by showing that it can handle data which lies on difficult manifolds, it can avoid problems caused by rough potential landscapes encountered by methods which solve PDEs directly, 
%and our method lends itself naturally to 
an adaptive sampling strategy that allows on-the-fly identification of and addition of data to regions important for predicting the statistics of interest.  Finally, we demonstrate that we can compute accurate statistics for a 75-dimensional model of sudden stratospheric warming.  This system provides a stringent test bed for our method.
\end{abstract}

%Immediately after the abstract, provide a maximum of 6 keywords, using American spelling and avoiding general and plural terms and multiple concepts (avoid, for example, 'and', 'of'). Be sparing with abbreviations: only abbreviations firmly established in the field may be eligible. These keywords will be used for indexing purposes.
% \begin{keyword}
% neural network, rare event, Feynman-Kac equation, high-dimensional PDE, adaptive sampling, Holton-Mass model
% %\MSC[2010] 00-01\sep  99-00
% \end{keyword}

% \end{frontmatter}

% \linenumbers

\maketitle

\section{Introduction}\label{sec:intro}
In many complex dynamical systems, behaviors of strong interest occur infrequently compared to the system's fastest timescale phenomena. For example, most climate-related destruction is due to extreme weather events (e.g., hurricanes, heat waves, flooding) \cite{Eastering2000climate,AghaKouchak2014global,Lesk2016influence,Mann2017influence,Frame2020attribution}. More broadly, fluid turbulence in both natural and engineered systems produces intermittent, damaging extreme events \cite{sapsis2021statistics}. In the molecular sciences, chemical reactions and molecular rearrangements occur on timescales many orders of magnitude longer than the timescale of individual bond vibrations \cite{Brooks1988,zwier2010reaching}.   In the biomedical sciences, it may take many mutations before a virulent strain of a pathogen emerges \cite{sohail2021mpl}, or many heart beats before a cardiac arrhythmia becomes life-threatening \cite{mirams2012application,liu2018mechanisms}.

Among the most common computational tasks related to these rare events is prediction---assessing the likelihood and extent of an event  (i.e., the risk and cost in the case of a deleterious event)---before it occurs.  When the event is not too rare, it can often be predicted with sufficient accuracy by direct forward-in-time integration of a computer model as is frequently done, for example, in weather prediction.  However, when the event is very rare, direct forward-in-time integration becomes prohibitively expensive because many simulated model trajectories are required to observe even one instance of the event, leave alone compute statistics. The computational cost increases further when the goal is to gain an understanding of how the rare event develops, which requires predictions generated from many initial conditions.

%These observations have motivated many attempts to address a basic question: Can limited simulations with an expensive, high-fidelity model be used to build forecasts of events that occur over long timescales?  
One common approach to this problem is to construct a ``coarse-grained'' model, in which some details of the system are treated implicitly \cite{marrink2007martini,saunders2013coarse,jumper_trajectory-based_2018}. One example is a Markov State Model (MSM), in which one groups the states of the full system into discrete sets and then evolves the system between these sets according to transition probabilities that are estimated from trajectories of the full system \cite{zwanzig_classical_1983,noe2009constructing,bowman2013introduction,husic_markov_2018}.   A variety of machine learning approaches that instead yield continuous coarse-grained representations of systems have come to be known as ``equation discovery'' \cite{kutz2017deep,rudy2017data,ogorman2018using,zanna2020data,chattopadhyay2020data,Kashinath2021mlweather}.  When an accurate coarse-grained model can be constructed, it can be simulated extensively to make predictions with statistical confidence.  However, building an accurate coarse-grained model can be challenging, in particular, because it is often not clear a priori which features must be included.  The construction of coarse-grained models thus remains a subject of intense inquiry. 

% An important goal in applied stochastic dynamics is to be able to make forecasts of a Markov process at a long time horizon without explicitly simulating extremely long trajectories.  In most systems of interest, the timescale for a given event of interest is much longer than what is available to a direct simulation.  One common approach is to use data to forecast the long time behavior of the system using short trajectories.  In  recent years, several machine learning approaches have been utilized to learn directly the dynamics of the system on either the full state space, or a reduced set of collective variables.  A large body of literature exists in the context of learning coarse-grained dynamics of molecular systems on reduced variables, so that the coarse grained system can be simulated for a long time. [cite cgnets, frank noe's recent stuff]  These approaches generally work by identifying a force field on some reduced set of variables which approximates the dynamics of the full system projected onto those reduced variables.  This learning is done either through linear least squares methods, or via more sophisticated neural network architectures.  These methods, while useful in principle, attempt to solve a very difficult problem since they ask for the full dynamical law.  Therefore when the system is very complex or the data availability is limited, they may not be practical.

Here, we pursue an alternative approach:  directly estimating conditional expectations of a Markov process as a function of initial condition.  We term these conditional expectations ``prediction functions.''  Prediction functions can be used to reveal how a rare event develops in remarkable detail.  For example, the committor (also known as the splitting probability)---the probability that a process proceeds to a set of target states before a competing set of states---can be used to define the transition state ensemble of a molecular rearrangement \cite{du1998transition,bolhuis2002transition}, as well as the pathways that lead between the reactant and product states \cite{strahan2021long,antoszewski2021kinetics}.  Prediction functions can also provide important information for decision making.  For example, the committor can be used by energy, transportation, and financial sectors to measure risk due to extreme weather and allocate resources accordingly \cite{Bloomfield2021subseasonal}. Committor estimation is a growing research focus in meteorology~\cite{Tantet2015early,Lucente2021coupling,Lucente2022committor,finkel2022revealing}. In real-time settings, the lead time---the expected time until onset of the event given that it occurs---is also essential to know \cite{finkel2021learning,finkel2021exploring,finkel2022revealing}.

%In this work we explore an alternative to equation discovery in which we use a data set of relatively short model simulations to train a neural network approximation of a rare event forecast {\it function}.

% forecast functions of the form
% \begin{equation}\label{eqn:genexp}
%   u(x) = \mathbbm{E}_x\left[g(X_T)e^{-\int_0^T V(X_s)ds}
%   + \int_0^T h(X_s)e^{-\int_0^s V(X_r)dr}\right] 
% \end{equation}
% where $X_t$ is the underlying dynamical system with $X_0=x$, the loss functions $g$, $V$, and $h$ are chosen by the user, and $T$ is the first escape time of $X$ from a domain. The rare event of interest is specified by the choice of domain. 
%  For example, by choosing $V=g=0$ and $h=1$ we compute the expected time until the rare event occurs. 

Prediction functions satisfy Feynman-Kac equations, linear equations of the operator that describes the evolution of expectations of functions of a process, the transition operator (also known as the Koopman operator \cite{williams_datadriven_2015}) and its infinitesimal generator \cite[Chapter~3]{pavliotis_stochastic_2014}.  Feynman-Kac equations cannot be solved by conventional discretization approaches because they involve a high-dimensional independent variable (the state of the underlying process).  Moreover, the form of the transition operator is generally not known.  Nonetheless, we showed recently that Feynman-Kac equations can be solved approximately by a basis expansion in which inner products of basis functions are estimated from a data set of short trajectories \cite{strahan2021long, thiede2019galerkin}.

While this approach has been successfully applied to such diverse processes as protein folding \cite{strahan2021long}, molecular dissociation \cite{antoszewski2021kinetics}, and sudden stratospheric warming \cite{finkel2021learning}, it relies on identifying an effective basis set.  One choice is to use a basis of indicator functions for discrete sets, in which case the approach reduces to construction of an MSM (but with appropriate boundary conditions for the prediction function).  However, just as it can be challenging to group states into sets that satisfy the Markov assumption in construction of an MSM \cite{husic_markov_2018,sidky_high-resolution_2019}, the choice of basis set is not always straightforward.  

Here, we address this issue through a neural network ansatz for prediction functions.  Our work builds on recent studies, which showed that a neural network ansatz can be used to solve for the committor if one assumes particular, explicit forms for the dynamical operator \cite{li2019computing,khoo2019solving,li_semigroup_2022,rotskoff2022active} (and see \cite{chen2021committor} for a closely related approach using tensor network approximation). Similar neural-network techniques have been devised to solve a wide variety of partial differential equations~\cite{han2018solving,han_solving_2020,karniadakis2021physics,raissi2019physics,chen2020friedrichs,zeng2022competitive}.
Because we work directly with a data set of short trajectories, our approach is free of restrictive assumptions about the dynamics { (e.g., microscopic reversibility)}
and does not require explicit knowledge of a model generating the data, opening the door to treating high-fidelity models, and even experimental and observational data \cite{finkel2022revealing}, without simplifying assumptions. 

%The basis expansion methods in \cite{strahan2021long, thiede2019galerkin} operationally build upon a collection of methods developed in the context of biomolecular simulation for the analysis of rare events \cite{bowman2013introduction}. 

%Nonetheless, as we demonstrate, forecast functions can be effectively approximated as solutions to the FKE by training a neural network representation on a data set of short trajectories of the dynamical system.

%There has been recent interest in solving high-dimensional partial differential equations using neural network approximations \cite{QVMC, Han}. 

In Section~\ref{sec:FK}, we review prediction functions and the Feynman-Kac equation that we need to solve to estimate them. 
In Section~\ref{sec:NNfinitelag} we introduce our neural network approach to solving Feynman-Kac equations using a data set of paired trajectories.
In Section~\ref{sec:NumericalConsiderations},
we compare with Galerkin methods and explore the role of the lag time and the distribution of trajectory initial conditions on performance.   In Section~\ref{sec:adapt} we introduce an adaptive sampling method that enriches the data set based on the current neural network approximation.  Finally, in Section~\ref{sec:HM} we apply our algorithm to estimating the probability of onset and the lead time of a sudden  stratospheric warming event.

\section{Prediction functions and their Feynman-Kac equations}\label{sec:FK}

We consider events defined by a set of target states $B$; often, there is also a competing set of states $A$.  For example, if we want to estimate the probability that a moderate storm develops into an intense hurricane before dissipating, we would take $B$ to include all weather states consistent with an intense hurricane and $A$ to include all quiescent states.  The initial moderate storm would be a state in the domain $D = (A\cup B)^c$.  Mathematically, we select states in $B$ with the indicator function
\begin{equation}
\mathbbm{1}_{B}(x) = \begin{cases}
1,& x\in B\\ 
0,& x\notin B,
\end{cases}
\end{equation}
where $x$ denotes a particular state of the system.  We define analogous indicator functions for other sets.

We assume the dynamics of the system can be described by a Markov process $X_t$.  In the example above, $X_0 = x$ is a moderate storm state, and the probability that it develops into an intense hurricane before the weather returns to a quiescent state is the committor:
\begin{equation}\label{eqn:qdef}
q(x)=\mathbb{P}_x[X_T \in B]=\mathbb{E}_x[\mathbbm{1}_B(X_T)],
\end{equation}
where the subscript indicates the initial condition, and $T=\inf\{t>0:X_t \in A\cup B\}$ is the stopping time, i.e., when the process leaves the domain $D = (A\cup B)^c$.

Continuing the example above, we may also want to compute the lead time, i.e., the average time until a moderate storm develops into an intense hurricane, given that the intense hurricane occurs ($B$ occurs before $A$).  The lead time tells us how much time we have to prepare for the worst case; by definition, it is shorter than the average time until an intense hurricane develops, which can be misleadingly large if the storm has a high probability of dissipating ($A$ occurs before $B$).  Mathematically, the lead time is
\begin{equation}
m_{AB}(x)=\frac{\mathbb{E}_x[T\mathbbm{1}_B(X_T)]}{\mathbb{E}_x[\mathbbm{1}_B(X_T)]}.
\end{equation}

When the event of interest is rare, computing $q(x)$ or $m_{AB}(x)$ by direct forward-in-time simulation is difficult.  It involves repeatedly simulating $X_t$ starting in a selected initial condition $x$ and running until either $A$ or $B$ is reached (which defines the stopping time $T$), and then assembling a sample average.  This approach has significant drawbacks: first, when the time $T$ is very large, generation of a single sample trajectory may be prohibitively computationally expensive, and second, when $q(x)$ is small, many sample trajectories will be required to observe a single trajectory reaching $B$.  For example, starting from a typical weather state, the expected time to the next extreme event may be years, and the probability that it occurs on a much shorter time scale may be very small.  %As a result, generating samples of $X_T$ may be prohibitively computationally expensive.  
%Second, when $q(x)$ is small, the standard error of estimates from independent trajectories, $\sqrt{q(x)}$, is large relative to $q(x)$ itself.  

In this paper we estimate prediction functions by solving operator equations for them approximately.  In the case of the committor, the operator equation takes the form
\begin{equation}\label{eqn:FKq}
(\mathcal{T}^{\tau}_{D^\textrm{c}}-{\cal I})[q](x)=0\ \text{with}\ q(x)=0\ \text{for}\ x\in A\ \text{and}\ q(x)=1\ \text{for}\ x\in B,
\end{equation}
where $\tau$ is a time interval known as the lag time and ${\cal I}$ is the identity operator. Here we focus on finite $\tau$; the case of infinitesimal $\tau$ is discussed in Section \ref{sec:lagtime}.  Above, the stopped transition operator $\mathcal{T}^s_{D^\textrm{c}}$ encodes the full dynamics of the system when it is in $D$; it is  defined by its action on an arbitrary test function $f$:
\begin{equation}\label{eqn:to}
\mathcal{T}^{\tau}_{D^\textrm{c}}[f](x) = 
\mathbb{E}_x\left[ f(X_{\tau\wedge T})\right],
\end{equation}
where $\tau\wedge T =\min\{\tau,T\}$.  Physically, \eqref{eqn:FKq} reflects the fact that the average probability that $B$ occurs before $A$ after time $\tau$ over all trajectories emanating from $X_0=x$ is the same as the probability that $B$ occurs before $A$ starting from $x$.
Similarly, the lead time satisfies
\begin{equation}\label{eqn:FKmab}
(\mathcal{T}^{\tau}_{D^\textrm{c}}-{\cal I})[m_{AB}q](x) =-\mathbb{E}_x\left[\int_0^{\tau\wedge T}q(X_s)ds\right] \ \text{with}\ m_{AB}=0\ \text{for}\ x\in A \cup B.
\end{equation}
In this case, the right hand side accumulates the time until reaching $A\cup B$, weighted by the likelihood of reaching $B$ before $A$.

Eqs.\ \eqref{eqn:FKq} and \eqref{eqn:FKmab} are examples of Feynman-Kac equations \cite[Chapter~3]{pavliotis_stochastic_2014}, which can take more general forms, such as
\begin{equation}\label{eqn:FK}
(\mathcal{T}^{\tau}_{D^\textrm{c}}-{\cal I})[u](x)    =
-\mathbb{E}_x\left[
\int_0^{\tau\wedge T} h(X_s)ds\right]\ \text{with}\ u(x)=g(x)\ \text{for}\ x\notin D
\end{equation}
which is solved by the prediction function.
\begin{equation}\label{eqn:genexp}
 u(x) = \mathbbm{E}_x\left[g(X_T) + \int_0^T h(X_s)ds\right] 
\end{equation}
We recover \eqref{eqn:FKq} by setting $h(x)=0$ and $g(x)=\mathbbm{1}_B(x)$ and \eqref{eqn:FKmab} by setting $h(x)=q(x)$ and $g(x)=0$; the latter case yields $[m_{AB}q](x)$, and we must solve separately for $q(x)$ and divide by it to obtain $m_{AB}(x)$.
Crucially, \eqref{eqn:FK} exactly characterizes $u$ for any choice of $\tau>0$.  In particular, $\tau$ can be chosen much shorter than typical values of $T$.

On its own, \eqref{eqn:FK} brings us no closer to a practically viable approximation of the prediction function.  The independent variable $x$ is typically high-dimensional, rendering useless any standard discretization approach to solving \eqref{eqn:FK} for $u$. Instead, the current state-of-the-art approach involves expansion of $u$ in a problem-dependent basis \cite{strahan2021long,antoszewski2021kinetics,thiede2019galerkin}.  In the next section, we explore a potentially more flexible and automated approach to solving $\eqref{eqn:FK}$.

\section{Solving Feynman-Kac equations with neural networks} \label{sec:NNfinitelag}

The goal of the present study is to solve \eqref{eqn:FK} by approximating $u$ by a neural network $u_\theta$ with a vector of parameters $\theta$.  Specifically, we seek $\theta=\theta^*$ that minimizes a mean square difference between the left and right hand sides of the Feynman-Kac equation and boundary condition in \eqref{eqn:FK}:
\begin{equation}\label{eqn:Min}
\theta^{*}=\argmin_{\theta}[C_\textrm{FKE} + \lambda C_{\rm BC}]
\end{equation}
with
\begin{equation}\label{eq:exactloss}
    C_\textrm{FKE} = \left\lVert \left((\mathcal{T}^{\tau}_{D^\textrm{c}}-{\cal I})u_{\theta}+ \mathbb{E}_x\left[\int_0^{\tau\wedge T}h(X_s)ds\right]\right)\mathbbm{1}_D\right\rVert_{\mu}^2\ \text{and}\ 
  C_{\rm BC} = \left\lVert (u_{\theta}-g)\mathbbm{1}_{D^c}\right\rVert_{\mu}^2.
\end{equation}
The norm that we use is the $\mu$-weighted $L^2$ norm $\lVert f\rVert_{\mu}^2 = \int \lvert f(x)\rvert^2 \mu(dx)$, where $\mu$ is the sampling distribution. 
% which we discuss in Section \ref{sec:sampling}.
{ 
Importantly, unlike the many existing estimators \cite{li2019computing,khoo2019solving,li_semigroup_2022,rotskoff2022active,chen2021committor,banushkina2015nonparametric,krivov2021blind,roux_string_2021,roux2022transition,e2005transition}, our data need not be generated from (or re-weighted according to) the invariant distribution of $X_t$ (which may not exist), a feature that we exploit in Section \ref{sec:sampling}. 
% Nor does the loss function in \eqref{eq:exactloss} require that $X_t$ be a reversible process. 
%Furthermore, we can treat  microscopically irreversible dynamics, as we do in Section \ref{sec:HM}.
% At first blush, \eqref{eq:exactloss} seems similar to others derived in the literature.  In particular, Roux recently proposed obtaining the committor by minimizing
% %\begin{equation}
% %\mathbb{E}_{\pi}\left[(q_+(X_0)-q_+(X_{dt}))^2\right]
% $\left\lVert \left(q_+(X_0)-q_+(X_{dt})\right)\right\rVert_{\pi}^2$
% %\end{equation}
% for some small lag time $dt$ \cite{roux_string_2021,roux2022transition} (see also \cite{banushkina2015nonparametric,krivov2021blind}).  Besides being limited to the committor, this loss function requires weighting initial conditions by the stationary distribution and microscopically reversible dynamics.  We show that better performance can be obtained from non-physical distributions in Section \ref{sec:sampling}, and we apply our method to a microscopically irreversible dynamics in Section \ref{sec:HM}.
}
In \eqref{eq:exactloss}, $C_\textrm{FKE}$ and $C_{\rm BC}$ are both zero when $u_\theta$ equals the desired prediction function.
The parameter $\lambda$ controls the strength of the first norm, which enforces the Feynman-Kac equation, relative to the second norm, which enforces the boundary condition.  Smaller values enforce the boundary conditions more strictly but can compromise the satisfaction of the Feynman-Kac equation. { For our numerical tests below, we tuned $\lambda$ by trial and error to the smallest value that still enforced the boundary conditions to the desired precision. }

{ 
The gradient of $C_\textrm{FKE}$ includes the integral of a product of two terms of the form $\mathcal{T}^{\tau}_{D^\textrm{c}}v$ with $v = u_\theta$ and $v = \partial_\theta u_\theta$, the gradient of $u_\theta$ with respect to the parameters $\theta$. While we cannot hope to evaluate $\mathcal{T}^{\tau}_{D^\textrm{c}}v$  exactly for any non-trivial $v$, as long as we can evaluate $v$ we have access to the random variable $v(X_{\tau \wedge T})$ whose expectation is $\mathcal{T}^{\tau}_{D^\textrm{c}}v$.  With only one sample of $X_{\tau \wedge T}$ for each sample of $X_0$, we would not be able to build an unbiased estimate of the product of two terms of the form $\mathcal{T}^{\tau}_{D^\textrm{c}}v$. One approach, common in reinforcement learning applications, is to simply drop the term involving this product from the gradient \cite{sutton2018reinforcement}. However, given at least two independent samples of $X_{\tau \wedge T}$ for each sample of $X_0$, we can construct an unbiased estimator of the full gradient of $C_\textrm{FKE}$ that converges to the exact gradient of $C_\textrm{FKE}$ in the limit of many samples of $X_0$ (even when the number of independent samples of $X_{\tau \wedge T}$ for each sample of $X_0$ does not increase). Below we outline a procedure that constructs an unbiased estimate of the gradient of $C_\textrm{FKE}$ given a data set of samples of $X_0$, together with $\ell\geq 2$ samples of $X_{\tau \wedge T}$ for each sample of $X_0$ (in tests of $2\leq \ell\leq 10$, we found the results to be insensitive to the choice of $\ell$, and we use $\ell=2$ throughout). } 

Our procedure is as follows.
\begin{enumerate}
    \item \label{step:initial} Select a set of $n$ initial conditions $X_0^{i}$ from the sampling distribution $\mu$.
    \item \label{step:trajs} From each $X_0^{i}$, launch $\ell$ independent unbiased simulations to generate trajectories $\{(X^{i}_{0},X^{i,j}_{\Delta},...,X^{i,j}_{S\Delta})\}_{j=1}^\ell$.  Here we assume that $\tau = S\Delta$.
\item For trajectory $j = 1, 2,\dots,\ell$ with initial condition $X_0^{i}$, determine the index of its stopping time as $k^{i,j}=\min\{s':X^{i,j}_{s'\Delta}\in(A\cup B)\text{ or }s'=S\}$. \label{step:stoppingtime}
\item \label{step:approxnorms} Given the data set of grouped trajectories, approximate the first norm in \eqref{eqn:Min} as
\begin{equation}\label{eq:CFKE}
\begin{split}
    \bar{C}_\textrm{FKE}= \frac{1}{n}\sum_{i=1}^n&\left(\frac{1}{|S_i|}\sum_{j\in S_i}u_{\theta}(X^{i,j}_{k^{i,j}\Delta})-u_{\theta}(X^{i}_{0})+\Delta\sum_{s=0}^{k^{i,j}}h(X^{i,j}_{s\Delta})\right) \\
    \times &\left(\frac{1}{|S'_i|}\sum_{j'\in S'_i}u_{\theta}(X^{i,j'}_{k^{i,j'}\Delta})-u_{\theta}(X^{i}_{0})+\Delta\sum_{s=0}^{k^{i,j'}}h(X^{i,j'}_{s\Delta})\right)\mathbbm{1}_{D}(X^{i}_0)
\end{split}
\end{equation}
and the second norm in \eqref{eqn:Min} as
\begin{equation}
   \bar{C}_{\rm BC}= \frac{1}{n}\sum_{i=1}^n \left (u_{\theta}(X^{i}_0)-g(X^{i}_0) \right)^2\mathbbm{1}_{D^c}(X^{i}_0).
\end{equation}
where $S_i$ and $S'_i$ are randomly chosen index sets such that $S_i \cap S'_i=\emptyset.$
\item Compute the total approximate loss function as
\begin{equation}\label{eq:approxcost}
    \bar{C} = \bar{C}_\textrm{FKE} + \lambda \bar{C}_{\rm BC},
\end{equation}
which converges to the loss in \eqref{eqn:Min} as $n$ increases. 
\item Adjust the parameters to minimize \eqref{eq:approxcost}.
\item Check termination criteria and stop if met (discussed further below). \label{step:stop} 
\item If adaptively sampling, apply the procedure in Section \ref{sec:adapt} and set $n$ to the total number of initial conditions.
\item Go to step \ref{step:stoppingtime}.
\end{enumerate}
% It may seem more natural to take $S_i=S'_i=\{1...l\}.$  Indeed this will guarantee that the loss function is positive.  In this case, however, the loss is only unbiased for large $l.$  
In principle, the loss can be minimized over any sufficiently flexible ansatz $u_\theta$.  In this work, $u_\theta$ is a fully connected feed-forward neural network, and we determine the optimal parameters via the 
Adam algorithm \cite{kingma_adam_2017}.  
{ 
In the present study, we stop training (step \ref{step:stop}) when the average loss for an epoch is less than zero. It is possible for $\bar{C}_{\textrm{FKE}}$ to become negative because  the two parenthetical factors in \eqref{eq:CFKE} are evaluated using independent samples of $X_{\tau \wedge T}$. 
While this could be avoided by choosing $S_i = S'_i$, the result would be a biased estimator of $C_\textrm{FKE}$. In the limit of large $n$, $\bar{C}_\textrm{FKE}$ converges to $C_\textrm{FKE}$, which must be non-negative.  When using any sample approximation of $C_\textrm{FKE}$, some regularization is required to avoid overfitting. 
% Using our unbiased loss function, however, with maximum overfitting, and a network constrained to give numbers between 0 and 1 for the committor, the loss function could assign 0.5 to each starting point, 0 to the first endpoint and 1 to the second endpoint, and the loss would then be $-0.25N$.  It is therefore necessary to introduce some form of regularization to avoid this.  
We find early stopping at the first occurrence of a negative value of  $\bar{C}_\textrm{FKE}$ to be a natural and  effective approach. 
}
Further details are given in conjunction with the numerical examples.

 %We refer to our method as the DYnamical Neural Operator Equation Solver (DYNOES).

\section{Illustration of numerical considerations}\label{sec:NumericalConsiderations}

In this section, we use a model for which we are able to compute reference results to illustrate the advantages of our approach relative to existing ones.  Specifically, we compare our approach with one that employs a basis expansion (a Markov State Model) and one that employs a neural network with an assumed form for the dynamical operator. Finally, we examine common choices for the sampling distribution.
{
We show that an important practical advantage of our approach is the freedom to choose the sampling distribution $\mu$ with which to weight the norm in $C_{\textrm{FKE}}$.
}

\subsection{M\"uller-Brown model}\label{sec:MB}

\begin{figure}[btp]
\begin{center}
\includegraphics[scale=0.6]{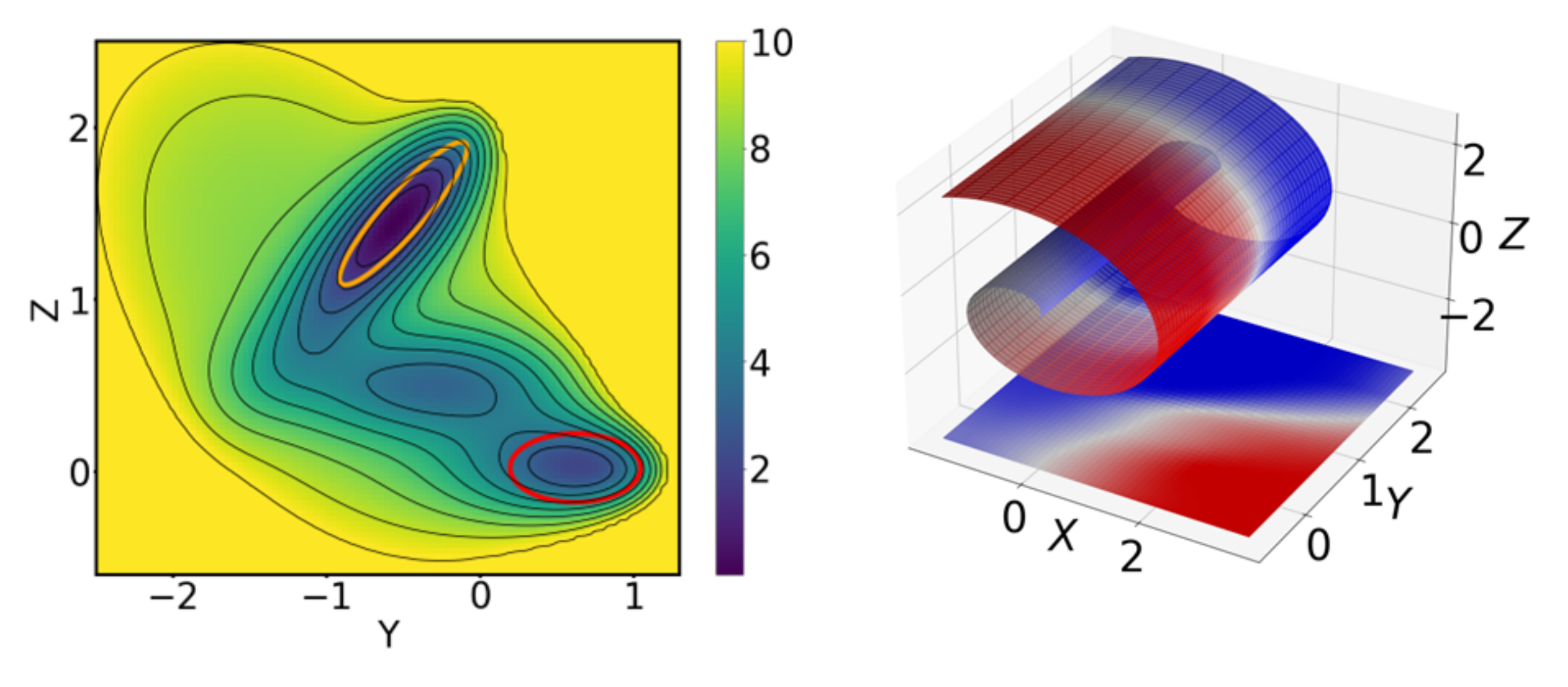}
\end{center}
\caption{\label{fig:Jelly}
The system used for the numerical experiments in Section \ref{sec:NumericalConsiderations}. (left) M\"uller-Brown potential \cite{muller_location_1979}.  Sets $A$ and $B$ are marked by the orange and red ellipses, respectively, and contours are spaced at intervals of 1 in the units of \eqref{eq:MB}.  (right)  Reference committor for the M\"uller-Brown dynamics mapped to the Swiss roll, and below on the two-dimensional surface.  We compute the reference from a finite difference scheme \cite{thiede2019galerkin} in two dimensions and then map it to the Swiss roll using \eqref{eq:Jelly}.
}
\end{figure}

The system that we study is specified  by the M\"uller-Brown potential \cite{muller_location_1979}, which is a sum of four Gaussian functions:
\begin{equation}\label{eq:MB}
V_{\rm MB}(y,z)=\frac{1}{20}\sum_{i=1}^4C_i \exp[a_i(y-z_i)^2+b_i(y-y_i)(z-z_i)+c_i(z-z_i)^2].
\end{equation}
For all results shown, we use $C_i=\{-200,-100,-170,15\}$, $a_i=\{-1,-1,-6.5,0.7\}$,
$b_i=\{0,0,11,0.6\}$, 
$c_i=\{-10,-10,-6.5,0.7\}$, 
$y_i=\{1,-0.27,-0.5,-1\}$, 
$z_i=\{0,0.5,1.5,1\}$.
The potential is shown in Figure \ref{fig:Jelly}(left). 

We consider the overdamped Langevin dynamics associated with $V_{\rm MB}$, discretized with the BAOAB algorithm \cite{leimkuhler_rational_2013}: 
\begin{equation}\label{eqn:MBEuler}
X_{t+dt}=X_t-\nabla V(X_t)dt+\sqrt{\frac{dt}{2\beta}}(Z_t+Z_{t-dt})
\end{equation}
where $dt$ is the time step, $\beta$ is the inverse temperature, $Z_t\sim N(0,1)$, $N(0,1)$ is the normal distribution with zero mean and unit standard deviation (i.e., $Z_t$ is Gaussian noise), and $V = V_{\rm MB}$ with $\beta=1$ unless otherwise specified.  In practice, we use a time step of $dt=0.001$, saving the configuration every time step, such that $\Delta=0.001$ (cf.\ step \ref{step:trajs} in Section \ref{sec:NNfinitelag}).  When the parameter $\beta$ is large, $X_t$ makes only very rare transitions between the local minima of $V_{\rm MB}$.

We define states $A$ and $B$ as
\begin{equation}\label{eqn:MBstates}
\begin{aligned}
&A=\{y,z:6.5(y+0.5)^2-11(y+0.5)(z-1.5)+6.5(z-1.5)^2<0.3\}\\
&B=\{y,z:(y-0.6)^2+5(z-0.02)^2<0.2\},
\end{aligned}
\end{equation}
neighborhoods of two of the three local minima of $V_{\rm MB}$ (Figure \ref{fig:Jelly}(left)).

The M\"uller-Brown model described above is commonly employed as a simple  illustration of the features of molecular rearrangements \cite{muller_location_1979,thiede2019galerkin, khoo2019solving,li_semigroup_2022,rotskoff2022active}.  The presence of local minima in addition to $A$ and $B$, and the fact that, at low noise, the trajectories connecting the minima do not align with the coordinate axes are both features that can be challenging for algorithms that enhance the sampling of transitions between the reactant ($A$) and product ($B$) states.  In our tests, we specifically focus on the committor.  We compute a reference committor by the finite difference scheme outlined in { the appendices of \cite{thiede2019galerkin,lorpaiboon2022augmented} with $\epsilon=0.0125$.  
% Specifically, we define the discrete approximation of the generator for Brownian dynamics using a grid of spacing $\epsilon=0.0125,$ and defining the matrix:
% $$
% P(y\pm \epsilon,z\pm \epsilon)=\frac{1}{1+e^{V(y\pm \epsilon,z\pm \epsilon)-V(y,z)}}
% $$
% We then solve for the reference committor as:
% $$
% diag(\mathbbm{1}_{(A\cup B)^c})(P-I)q_+=-diag(\mathbbm{1}_{(A\cup B)^c})(P-I)\mathbbm{1}_{B}
% $$
In all tests, we compare the estimated committor to the reference committor computed using the same potential energy function used to generate the data.}

Finally, to represent the fact that one of the most challenging aspects of treating complex systems is that the manifold on which the dynamics take place is generally not known, we transform the trajectories and sets $A$ and $B$ to a new set of coordinates.  
Specifically, we map the two-dimensional system onto a Swiss roll (Figure \ref{fig:Jelly}(right)):
  \begin{align}\label{eq:Jelly}
         \begin{bmatrix}
           y \\
           z \\
         \end{bmatrix}
         \to
         \begin{bmatrix}
           (c+y) \cos((y+c)f) \\
           z  \\
           (c+y) \sin((y+c)f)
         \end{bmatrix},
  \end{align}
where the parameter $f$ controls how tightly the roll is wound, and $c$ is an offset to ensure that the range of $x$ is positive.  Unless otherwise specified, we use $f=3$ and $c=1.8$.  
{
For the remainder of the tests based on the M\"uller-Brown potential, we use the three-dimensional coordinates as input features for all neural networks and $k$-means clustering.  For clarity of visualization, we plot the estimated committors on the original two-dimensional coordinates. The error metric that we use is independent of coordinate system.
}

{
Unless otherwise specified, for our experiments with the M\"uller-Brown model below, we draw 30,000 initial conditions} uniformly from the region:
\begin{equation}\label{eq:initregion}
\Omega=\{y,z:-1.5<y<1.0,-0.5<z<2.5,V(y,z)<100\}.
\end{equation}
Two independent trajectories of length $\tau$ (to be specified below) are then generated from each initial condition using \eqref{eqn:MBEuler}.

\subsection{Neural network details}\label{sec:netdetails}

For all the numerical experiments involving the M\"uller-Brown potential, we use fully connected feed-forward neural networks with three inputs, three hidden layers, each consisting of 30 sigmoid activation functions, and an output layer with a single sigmoid activation function.  
In all trials with fixed data sets, we trained for a maximum of 3000 epochs with a learning rate of 0.0005 and a batch size of 1500.  Each epoch proceeds by drawing a permutation of the data set, then one step of Adam is performed using mini-batches of size 1500 (that is, 1500 pairs of trajectories) such that each trajectory pair is used exactly once per epoch 
{
(that is, the number of Adam steps is the data set size divided by the mini-batch size).  The boundary term is computed with the same mini-batch as $\bar{C}_\textrm{FKE}$; we use $\lambda=1$ to weight the terms in the loss function.} %
We also explored deeper networks with ReLU activation functions, and they performed comparably and generally required shorter training times (results not shown); we focus on the shallower networks with sigmoid activation functions because they allow a direct comparison with loss functions involving explicit derivatives of $u_\theta$ in Section \ref{sec:lagtime}.  

\subsection{Galerkin methods}\label{sec:SR}

As discussed in the Introduction, one of our main motivations in introducing an approach based on neural networks is that it can be difficult to identify basis functions for linear (e.g., Finite Element or other Galerkin) methods for solving Feynman-Kac equations.  To illustrate this issue explicitly, we compare estimates for the committor from our approach with those obtained from dynamical Galerkin approximation \cite{thiede2019galerkin,strahan2021long} using a basis of indicator functions, which can be considered a MSM \cite{thiede2019galerkin}.  We do so as a function of the parameter $f$ in \eqref{eq:Jelly} and generate data sets with $0\leq f\leq 10$.

To construct an MSM, we clustered the configurations in each data set by the $k$-means algorithm (with $k$ as specified below) applied to the three-dimensional coordinates of the model. The indicator functions of the set of points closest to each cluster centroid form a basis for a Galerkin approximation of the committor function.  A transition matrix ${\bf T}$ was constructed by counting transitions of the stopped process between clusters among our trajectory data set with  a lag time of $\tau=150\Delta$.
% from $X_0^{i}$ and $X_{\tau\wedge T}^{i,1}$ using a lag time of $\tau=150\Delta$. 
Here we use the convention that the row and column indices are zero for $A$ and their maximum values for $B$.  The committor is then computed from $\bf Tq_+=q_+$ with the last component of the solution vector set to 1.
The neural network and its training were as described in Section \ref{sec:netdetails}.
%We use fully connected feed-forward neural networks with three inputs, three hidden layers, each consisting of 30 sigmoid activation functions, and an output layer with a single sigmoid activation function.  
%For each trial, we trained for a maximum of 3000 epochs with a learning rate of 0.0005 and a batch size of 1500.  Each epoch proceeds by drawing a permutation of the data set, then one step of Adam is performed using mini-batches of size 1500 (that is, 1500 pairs of trajectories) such that each trajectory pair is used exactly once per epoch.  

%On the left, we compare the network trained with a lag time of 20 and 100000 trajectory pairs with a MSM with 300 clusters and the same data set.

\begin{figure}[btp]
\includegraphics[scale=0.6]{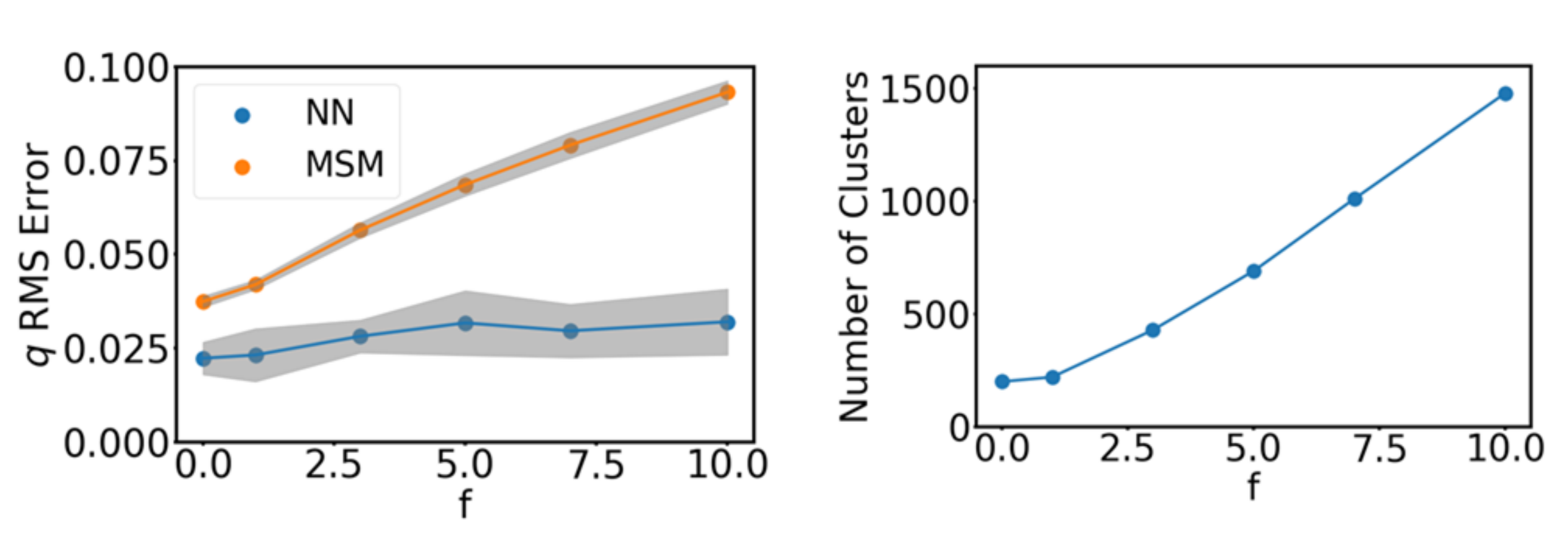}
\caption{\label{fig:Jelly_Results}
Comparison with Galerkin methods. (left) For an MSM estimate with { $k=300$ clusters}, the root mean square (RMS) error in the committor for the M\"uller-Brown model as the Swiss roll is wound tighter (higher $f$ in \eqref{eq:Jelly}). 
Shading shows the standard deviation in the error from training on ten independent data sets.
(right) Number of MSM clusters needed to achieve an RMS error in the committor of less than 0.045. 
}
\end{figure}

Figure \ref{fig:Jelly_Results} shows the results.  
We see that as the roll is wound tighter (higher $f$), the MSM estimates, {constructed with a constant 300 clusters,} decrease in accuracy, while the network estimates remain consistently good.  In the right panel, we vary the number of clusters and report the number required to reach a root mean squared error threshold of 0.045. This threshold is chosen because it results in  numbers of clusters in a range that is typical in MSM studies~\cite{strahan2021long,finkel2021learning}.   We increase the number of trajectories in proportion to the number of MSM clusters to ensure that each cluster is sampled a consistent amount.  In this test, we see that large numbers of MSM clusters, and hence large amounts of data, are needed.  Intuitively, the MSM encounters problems when a single cluster spans adjacent layers.  Therefore, it is necessary to vary the size of the clusters with the distance between layers of the Swiss roll, which is a linear function of $1/f$.  Consistent with this idea, in Figure \ref{fig:Jelly_Results} we find an approximately linear dependence of the number of clusters needed to achieve a certain error threshold. 

We note that in practice MSMs are often constructed on coordinates obtained from a method for dimensionality reduction and/or manifold learning.  With such pre-processing, linear methods can clearly be successful.  However, kernel-based methods for dimensionality reduction (e.g., diffusion maps \cite{coifman_diffusion_2006} or kernel time-lagged independent component analysis \cite{schwantes_modeling_2015,bittracher_dimensionality_2020}) scale poorly with the size of the data set.
A neural network (e.g., an autoencoder \cite{kingma_auto-encoding_2014,wehmeyer_time-lagged_2018}) can be used for dimensionality reduction, but the approach presented here is simpler in that we go directly from model coordinates to prediction function estimates.

\subsection{Lag time}
\label{sec:lagtime}

As discussed in the Introduction, neural networks have been applied to estimating high-dimensional committors assuming a partial-differential form for the dynamical operator \cite{li2019computing,khoo2019solving,li_semigroup_2022}.  This form arises in the limit that one considers an infinitesimal lag time.  In this case, one can write 
\eqref{eqn:FK} as
\begin{equation}\label{eqn:FK3}
\mathcal{L}[u](x)=h(x)\ \textrm{for}\ x\in D\ \textrm{and}\ 
u(x)= g(x)\ \text{for}\ x\notin D,
\end{equation}
where $\mathcal{L}$ is the is the infinitesimal generator:
\begin{equation}\label{eqn:gen}
\mathcal{L}[f](x) = \lim_{dt\rightarrow 0}\frac{\mathbb{E}_x\left[f(X_{dt})\right]-f(x)}{dt}.
\end{equation}
For a diffusion process, $\mathcal{L}$ takes the form
\begin{equation}\label{eq:Ldef}
    \mathcal{L}f(x)=\sum_{i=1}^{\kappa}b_i(x)\frac{\partial f}{\partial x_i} +
    \frac{1}{2}\sum_{i,j=1}^{\kappa}(\sigma \sigma^\textrm{T})_{ij}(x)\frac{\partial^2 f}{\partial x_i\partial x_j},
\end{equation}
where $b\in \mathbb{R}^{\kappa}$ and $\sigma\in \mathbb{R}^{\kappa\times \kappa}$ are the drift and diffusion coefficients that determine the evolution of $X_t$.   In the limit of small $dt$,  the dynamics in \eqref{eqn:MBEuler} correspond to a generator with $b = -\nabla V$ and $\sigma = \sqrt{2/\beta}$.  In this case, the loss function becomes
\begin{equation}
\bar{C}_\textrm{FKE}=\label{eq:0LagLoss}
    \sum_{i=1}^n\left(\sum_{j=1}^{\kappa}b_j(X_0^i)\frac{\partial u_{\theta}(X_0^i)}{\partial x_j} +
    \frac{1}{2}\sum_{j,l=1}^{\kappa}(\sigma \sigma^\textrm{T})_{jl}(X_0^i)\frac{\partial^2 u_{\theta}(X_0^i)}{\partial x_j\partial x_l}-h(X_0^i)\right)^2\mathbbm{1}_{D}(X_0^i)
\end{equation}
with an appropriate boundary condition term.  
% {
% Infinitesimal lag times are also assumed in deriving the estimator $\left\lVert \left(q_+(X_\tau)-q_+(X_0)\right)\right\rVert_{\pi}^2$ from a reactive flux expression in \cite{banushkina2015nonparametric,krivov2021blind,roux_string_2021,roux2022transition} despite its superficial similarity to \eqref{eq:exactloss}; in this sense, it is closely related to $\left\lVert \nabla q\right\rVert_{\pi}^2$ in \cite{li2019computing,khoo2019solving,li_semigroup_2022}, which derives from \eqref{eq:0LagLoss} \cite{e2005transition}.
% }

{
The loss function in \eqref{eq:0LagLoss} differs from the one used in many recent articles on the subject of committor estimation with neural-network (or recently tensor-network) approximations \cite{li2019computing,khoo2019solving,li_semigroup_2022,rotskoff2022active,chen2021committor}. Those papers focus specifically on the case of reversible overdamped diffusive dynamics. In this case the committor can be found by minimizing a sample approximation of the loss function
$\left\lVert \nabla q\right\rVert^2_\pi$
(for constant, isotropic diffusion coefficient)
where $\pi$ is the invariant distribution of the dynamics \cite{e2005transition}.  Relatedly, despite a resemblance to \eqref{eq:exactloss}, the estimator $\left\lVert \left(q_+(X_\tau)-q_+(X_0)\right)\right\rVert_{\pi}^2$ that appears in \cite{banushkina2015nonparametric,krivov2021blind,roux_string_2021,roux2022transition} is, in fact, a small $\tau$ approximation of $\left\lVert \nabla q\right\rVert^2_\pi$.
}

We stress that \eqref{eq:0LagLoss} is only appropriate for diffusion processes and requires working with the full set of variables in which the dynamics are formulated.  Importantly, one generally analyzes only functions of a subset of the variables (termed collective variables or order parameters)~\cite{strahan2021long, antoszewski2021kinetics, finkel2022revealing, thiede2019galerkin}.  For example, in a molecular simulation of a solute in solvent, one may include only the dihedral angles of the solute.  In a weather simulation, one may focus on the wind speed and geopotential height at particular altitudes.  When working with observational data, one only has access to the features that were measured.  Even when the tracked variables can be described by an accurate coarse-grained model, that model is not known explicitly and is difficult to identify from data. %In other words, we rarely know $b$ and $\sigma$.
These considerations make minimization of any loss function explicitly involving \eqref{eq:Ldef} impossible for many practical applications.  

%Despite the fact that \eqref{eq:Ldef} is thus at best an approximation in all practical settings, 
Nonetheless, the loss function in \eqref{eq:0LagLoss} is appealing because it involves only a single time point,  so no trajectories need to be generated if explicit forms for the drift and diffusion coefficients are known.  While this would appear to be an advantage, we show in this section that, even when the dynamics can be reasonably described by \eqref{eq:Ldef}, it can be preferable to work with finite lag times.

\begin{figure}[hbt]
\begin{center}
\includegraphics[scale=0.6]{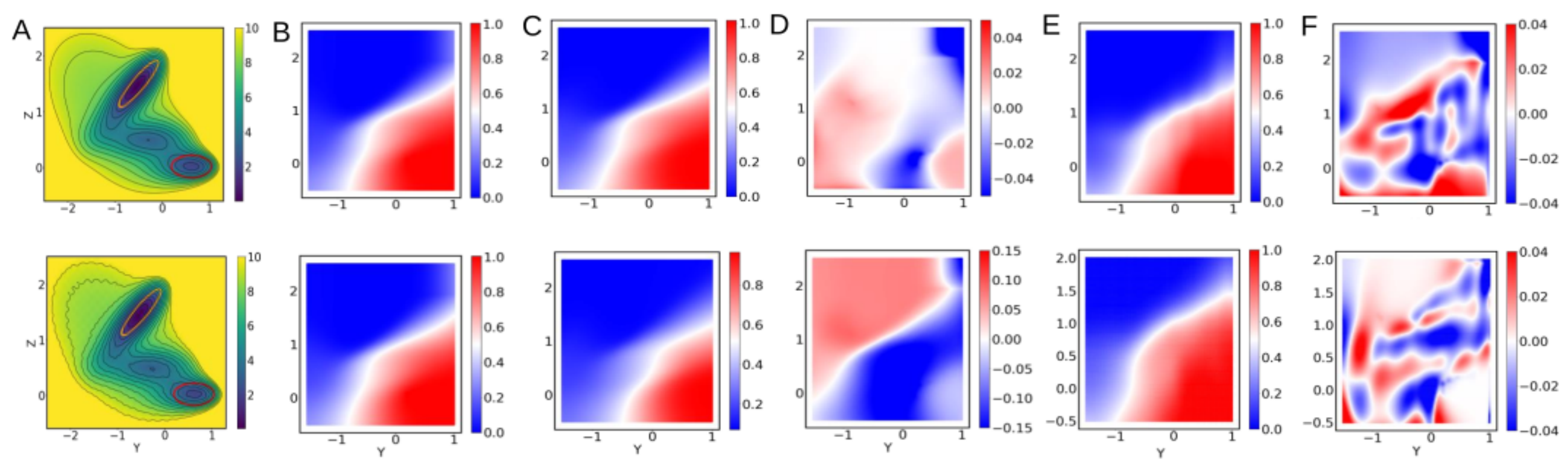}
\end{center}
\caption{\label{fig:MB_pics}
Effect of potential roughness on the performance when using a loss function based on the infinitesimal lag time limit.  Results shown are obtained with \eqref{eq:0LagLoss} and \eqref{eq:roughMB} with $\omega=0$ (top row), and $\omega=10$ (bottom row).
(A) The potentials.  (B) Reference committors obtained from the finite-difference scheme in \cite{thiede2019galerkin, lorpaiboon2022augmented}.  (C) Neural network prediction of the committors.  (D) Differences between the references and the predictions. 
{
(E,F) Same as columns C and D, except for a lag time of 100 steps.  Note the different scales on the difference maps in columns D and F.}
}
\end{figure}

To make this point, we consider dynamics governed by the M\"uller-Brown potential with a small oscillating perturbation (Figure \ref{fig:MB_pics}A):
\begin{equation}\label{eq:roughMB}
V(x)=V_{\rm MB}(x)+0.1\sin(2\pi \omega x)\sin(2\pi \omega y),
\end{equation}
where $\omega$ controls the spatial frequency of the perturbation.
Again we represent the data on the Swiss roll as described in Section \ref{sec:MB}.  As shown in Figure \ref{fig:MB_pics}B, the perturbation is sufficiently small that it makes no qualitative change to the committor.   

Given this data set, we train neural networks to minimize the loss function in \eqref{eq:approxcost}, using either \eqref{eq:CFKE} or \eqref{eq:0LagLoss} for $\bar{C}_\textrm{FKE}$, with $h(x) = 0$ and $g(x) = \mathbbm{1}_B(x)$, corresponding to the committor.  The network architecture was the same as above:  i.e., fully connected feed forward with two inputs, 30 activation functions per hidden layer, and one output. The neural network and its training were as described in Section \ref{sec:netdetails}.
% All networks were trained for 3000 epochs with a learning rate of 0.0005 and a batch size of 1500.  
%For the infinitesimal lag time method, we found that it was necessary to use sigmoid activation functions owing to the derivatives of $u_\theta$ in the loss function.  Because the sigmoid activation functions have derivatives that are less than one, we found that we needed a shallower network (we use three hidden layers for the results shown) when using the sigmoid activation functions.  Otherwise the network architecture was the same as above:  i.e., fully connected feed forward with two inputs, 30 activation functions per hidden layer, and one output.  The finite lag time networks were the same as above:  i.e., with ten hidden layers and ReLU activation functions.  Again all networks were trained for 1000 epochs with a learning rate of 0.0005 and a batch size of 1500.  

%\begin{figure}[hbt]
%\begin{center}
%\includegraphics[scale=0.6]{Figures/MBRefs2.pdf}
%\end{center}
%\caption{\label{fig:MB_pics2}
%Similar test to figure \ref{fig:MB_pics}, except %this time using a lag time of 100 timesteps.
%}
%\end{figure}

\begin{figure}[hbt]
\begin{center}
\includegraphics[scale=0.6]{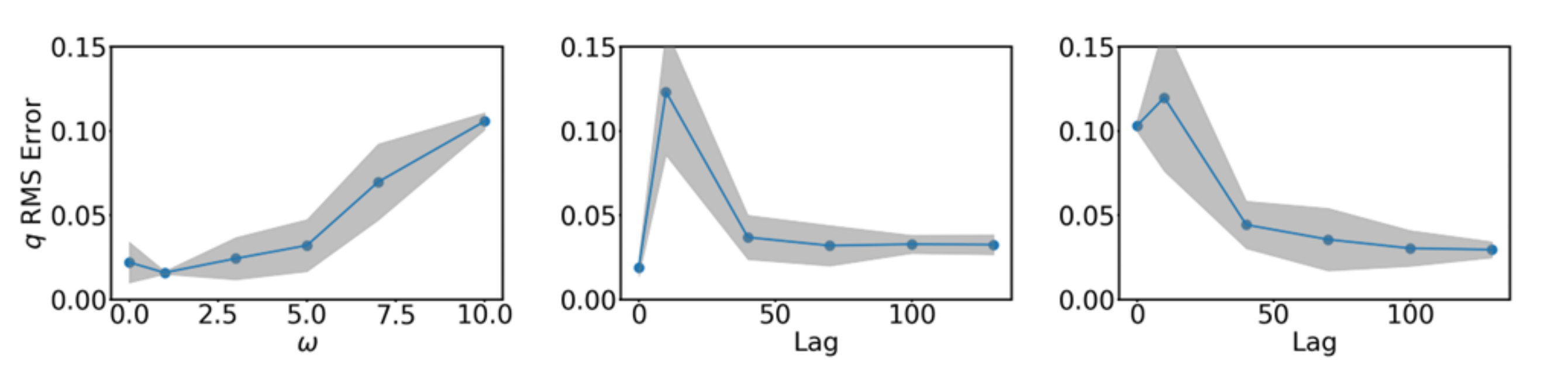}
\end{center}
\caption{\label{fig:fig:finitelagresults}
Comparison of infinitesimal and finite lag time loss functions.  (left) RMS error in the committor for the M\"uller-Brown dynamics mapped to the Swiss roll obtained with the infinitesimal lag time loss function in \eqref{eq:0LagLoss} as the frequency of the sinusoidal perturbation, $\omega$, is increased.
The other panels show the error as the lag time is increased with the frequency fixed at $\omega=0$ (center) or $\omega=10$ (right). Shading shows the standard deviation in the error from training on ten independent data sets.
}
\end{figure}

Typical results are shown in Figure \ref{fig:MB_pics}C and D, and the error in the committor is quantified in Figure \ref{fig:fig:finitelagresults}.
As the frequency of the the perturbation increases, the drift becomes large, with rapid sign changes, and the training of the infinitesimal lag time network tends to get stuck at poor estimates of the committor (Figure \ref{fig:fig:finitelagresults}(left)).  By contrast, finite lag time networks consistently achieve low errors at longer lag times (Figure \ref{fig:fig:finitelagresults}(center and right)).  This presumably results from averaging over values of the drift.
Interestingly, we found that when the potential is smooth (Figure \ref{fig:fig:finitelagresults}(center)), slightly lower errors can be obtained using the zero lag time approach.  However, in the presence of even such a small amount of roughness that the committor is qualitatively unchanged (Figure \ref{fig:MB_pics}A and B), our finite lag time approach performs better (Figure \ref{fig:fig:finitelagresults}(right)).  We expect the latter case to be more relevant in many practical applications.  
%{
%We note that the estimator $\left\lVert \left(q_+(X_0)-q_+(X_{\tau})\right)\right\rVert_{\pi}^2$ in \cite{banushkina2015nonparametric,krivov2021blind,roux_string_2021,roux2022transition} appears to use finite lag times, but it is derived through an infinitesimal lag time expression for the reactive flux, and we thus expect it to perform similarly to \eqref{eq:0LagLoss} for small $\tau$.
%}

It may be tempting to assume that the zero lag time approach has lower computational cost since there is no need to actually simulate the stochastic differential equation (here, \eqref{eqn:MBEuler}).  This is not necessarily the case.  With the infinitesimal lag time loss function, the drift needs to be evaluated for each data point for every pass over the data set (one epoch).  By contrast, the finite lag time loss function introduced here does not require evaluation of the drift once the data set is generated.
%This is the same cost as simulating one time step of \eqref{eqn:MBEuler} for each initial condition.  
Therefore, if the number of epochs needed to train the zero lag time network is comparable to the number of time steps used to generate the data set for the two-trajectory method, the finite lag time method will be less computationally costly.

\subsection{Sampling distribution}
\label{sec:sampling}

\begin{figure}[btp]
\begin{center}
\includegraphics[scale=0.4]{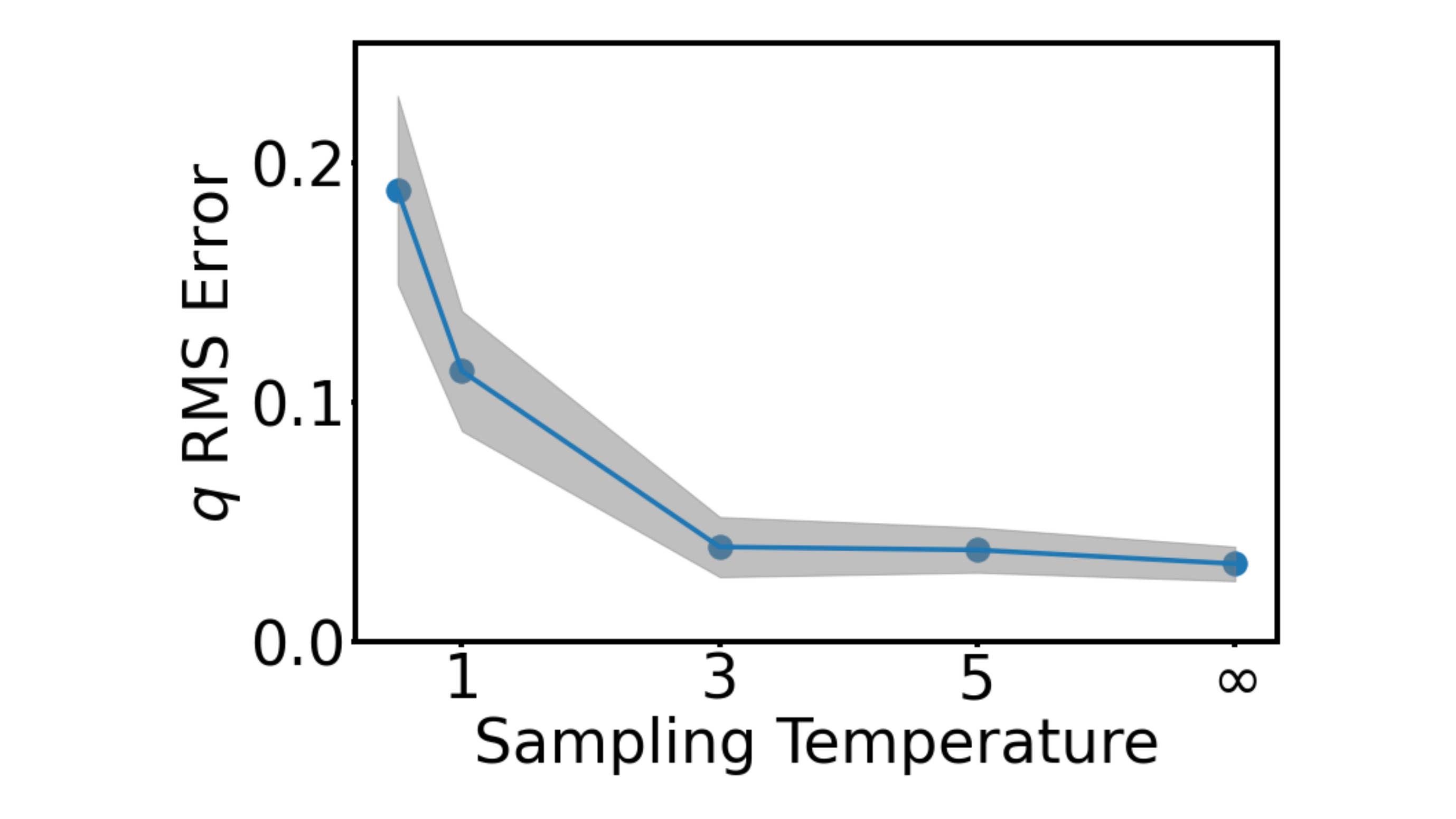}
\end{center}
\caption{\label{fig:MB_QV}
RMS error of the committor as the sampling temperature, $1/\beta_s$, is increased. The point at $1/\beta_s = 1$ corresponds to the stationary distribution for the M\"uller-Brown model (in the small $dt$ limit) at the temperature of the dynamics.  Increasing the temperature makes the distribution more uniform.  The point labeled $\infty$ is uniform.  Shading shows the standard deviation in the error from training on ten independent data sets.
}
\end{figure}

In this section, we investigate the role of the choice of sampling distribution.  
Following generation of a data set as described in Section \ref{sec:MB}, we selected initial points from the region specified in \eqref{eq:initregion} with weight $\mu(x)\propto\exp(-\beta_sV(x))$ ($\beta_s$ need not be the same as $\beta$) and trained over a range of $\beta_s$ values. 
To a good approximation when $dt$ is small, the invariant density of \eqref{eqn:MBEuler} is $\mu$ with $\beta_s=\beta$.  When $\beta_s$ is large, the data set of initial conditions concentrates at the local minima of $V(x)$.  As $\beta_s$ becomes small, the distribution approaches uniform.  %We will explore the effect of this choice on training accuracy.  
While this parametric form for the sampling distribution is convenient for the tests performed in this section, we emphasize that,
unlike many existing schemes \cite{li2019computing,khoo2019solving,li_semigroup_2022,roux_string_2021,roux2022transition}, our algorithm does not require explicit knowledge of the invariant density.

%The network architecture and training procedure were the same as the previous finite lag time examples. 
We trained ten networks on each pair of sampling distribution and lag time following the procedure in Section \ref{sec:netdetails}, and the resulting errors and their standard deviations are plotted as a function of $1/\beta_s$ in Figure \ref{fig:MB_QV}.
At low $1/\beta_s$, which concentrates the initial points in the minima, the network is unable to find a good solution at any lag time.  As $1/\beta_s$ increases and the distribution becomes more uniform, the solution improves significantly. This suggests that it is important to have the regions between minima well-represented in the data set, which is consistent with previous observations \cite{ma_automatic_2005,thiede2019galerkin,strahan2021long,finkel2021learning, rotskoff2022active}.  In high-dimensional examples, sampling the transition regions is not straightforward, and we present a solution to this problem in the next section.

\section{Adaptive Sampling}\label{sec:adapt}
As we showed in Section~\ref{sec:sampling}, the choice of sampling distribution is important.  
%We noted that it is important to concentrate sampling in the transition regions, rather than in the basins. 
In this section, we propose a simple method for adding data 
%to transition regions 
as the training proceeds. 
%More specifically, we add data points proportional to the residual squared of ... projected onto a space of collective variables $\psi$.  Many enhanced sampling methods require biasing the dynamics along collective variables which separate the system's metastable states (i.e., umbrella sampling or metadynamics \cite{barducci_well-tempered_2008,kumar_weighted_1992}). Here, we adopt a similar strategy. 
Since the approach depends on constructing a spatial grid we must first select a low-dimensional (e.g., two-dimensional) set of (possibly non-linear) coordinates $\xi(x)$ which, as noted above, we term collective variables.
We then partition the space of possible $\xi$ values
 into bins of equal volume  labeled $S_1,S_2,\dots,S_m$, and estimate
\begin{equation}\label{eq:adaptive}
    P_\eta=\frac{\int\left((\mathcal{T}^{\tau}_{D^\textrm{c}}-{\cal I})u_{\theta}+\mathbb{E}_x\left[\int_0^{\tau\wedge T}h(X_s)ds\right]\right)^2\mathbbm{1}_{S_\eta}(\xi(x))\mu(x)dx}{\int\mathbbm{1}_{S_\eta}
    (\xi(x))\mu(x)dx}
\end{equation}
for each bin.  The weights $P_\eta$ are then used to select bins, and new initial points are then sampled from the selected bins with uniform probability.
The essential idea is that we  add data to the regions (bins) that contribute most to $C_\textrm{FKE}$.

When adaptively sampling to learn the committor we approximate \eqref{eq:adaptive} by
\begin{equation}\label{eq:adaptiveData}
\hat{P}_\eta =\frac{\sum_{i=1}^n\left(\frac{1}{|S_i|}\sum_{j\in S_i}(u_{\theta}(X^{i,j}_{k^{i,j}})-u_{\theta}(X^{i,j}_0))\right)\left(\frac{1}{|S'_i|}\sum_{j\in S'_i}(u_{\theta}(X^{i,j}_{k^{i,j}})-u_{\theta}(X^{i,j}_0))\right)\mathbbm{1}_{S_\eta}(\xi(X^{i,1}_0))}{\sum_{i=1}^n\mathbbm{1}_{S_\eta}(\xi(X^{i,1}_0))}.
\end{equation}
We compute \eqref{eq:adaptiveData} for each bin, select $N$ bins with probability proportional to $\hat{P}_\eta$ with replacement, and sample a single additional initial point  from each selected bin.  From each of the $N$ new initial points we generate a trajectory. 
{
In practice, we observed that  \eqref{eq:adaptiveData} can become negative for some  bins in the same way that $\bar{C}_\textrm{FKE}$ can become negative.  In this case, we set all negative probabilities to zero.  
}
%Intuitively, the numerator of \eqref{eq:adaptiveData} tends to be largest where the committor is steepest, which is at $q\approx 0.5$; \eqref{eq:adaptiveData} thus tends to drive sampling to the transition region.

% Clearly the success of the method depends on the choice of $\psi$.  We expect the best $\psi$ to be ones that can separate the transition region from states $A$ and $B$, so as to ensure that the transition region is adequately sampled and the network can resolve the dynamics in it.  Capturing the essential features of the dynamics in the transition region is important because the statistics there generally dominate the statistics of the overall rare-event process. We note that, while identifying transition regions of complex systems is notoriously hard, it is often possible to find collective variables that are sufficient to separate the transition region from states $A$ and $B$.  This is the idea behind many methods for enhancing the sampling of rare events \cite{henin2022enhanced,dickson2010enhanced,zwier2010reaching}.

The success of our adaptive sampling approach depends on the choice of $\xi$.  In the absence of other knowledge, a reasonable choice is the current estimate of the committor function itself. 
%This choice for the collective variable was also made for an importance sampling scheme combining umbrella sampling \cite{thiede2016eigenvector,dinner2020stratification} and replica exchange \cite{swendsen1986replica,fukunishi2002hamiltonian} for learning rare events recently \cite{rotskoff2022active}. 
% This choice was taken in \cite{rotskoff2022active}, which generated new training data points for minimization of a cost function similar to \eqref{eq:0LagLoss} using an umbrella sampling algorithm \cite{thiede2016eigenvector,dinner2020stratification} that discretized space according to the values of the learned committor approximation. 
We adopt this choice  to test our adaptive sampling procedure on the M\"uller-Brown model.  A related adaptive sampling approach using stratified sampling~\cite{thiede2016eigenvector,dinner2020stratification} based on a current committor estimate is proposed in~\cite{rotskoff2022active}.

\begin{figure}[btp]
\begin{center}
\includegraphics[scale=0.5]{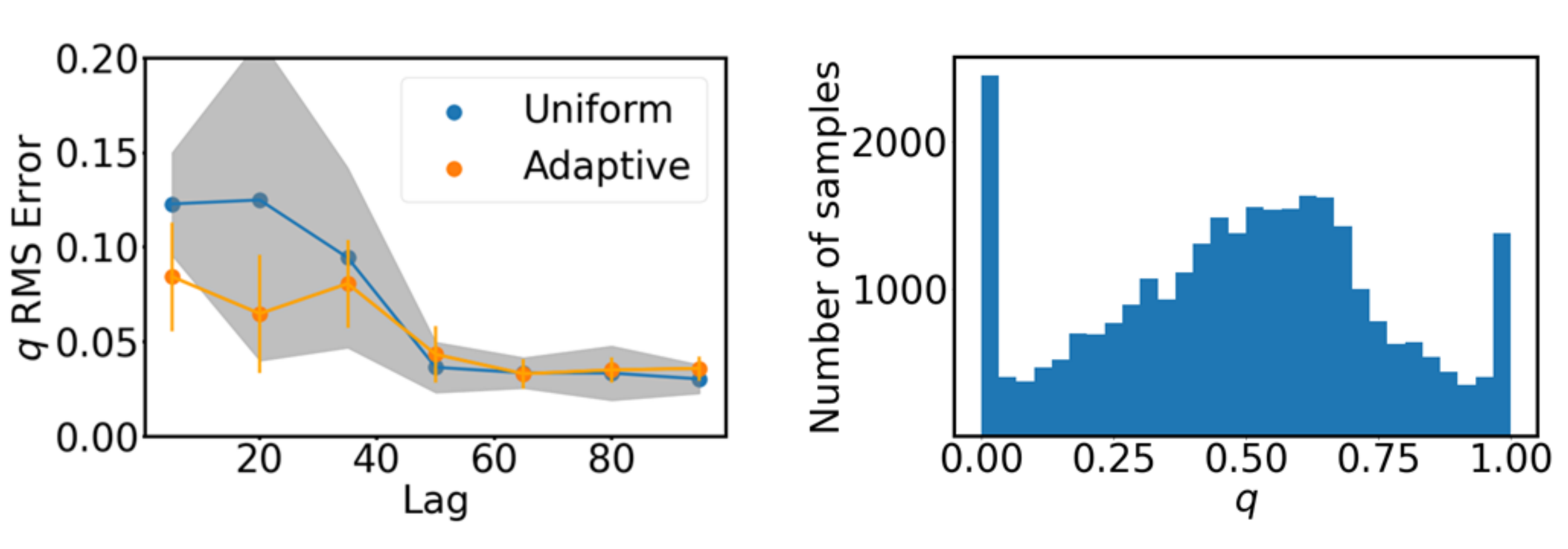}
\end{center}
\caption{\label{fig:Adaptive}
Adaptive sampling scheme applied to the M\"uller-Brown dynamics mapped to the Swiss roll. (left) Comparison of uniform and adaptive sampling.  Shading and error bars show the standard deviation in the error from training on ten independent data sets. (right) Histogram of the final data set as a function of the committor from training a neural network with a lag time of 100.}
\end{figure}

The simulation and Swiss roll parameters, as well as neural network and training parameters are the same as above.  We initially train with 10,000 pairs of { trajectories drawn uniformly from the region in \eqref{eq:initregion} for 1000 epochs. Then we alternate between adding $N=5000$ new pairs of trajectories and training for 500 epochs, for four cycles.  We compare to 30,000 trajectory pairs drawn uniformly from the region in \eqref{eq:initregion}.  Results are presented in Figure \ref{fig:Adaptive}.}  We find that the adaptive sampling and uniform sampling perform similarly at long lag times, although the adaptive procedure gives more reproducible results as shown by the smaller error bars. At short lag times the average error is lower as well.  { The adaptive sampling procedure concentrates sampling in the transition region, that is, near $q_+=0.5.$  In the next section, we illustrate the adaptive sampled distribution on our atmospheric model, and we again see that sampling is effectively directed to the transition region.  }{  For low noise diffusions, the transition region becomes narrower, and this is reflected by a sharper peak than in Figure \ref{fig:Adaptive}.  In our testing, our adaptive sampling scheme remains effective, although more data are required at lower noise.  We find that our method works for barriers $<10/\beta$.} 

% We illustrate the adaptive sampling procedure on the M\"uller-Brown potential.  Since we do not have convenient collective variables for the dynamics on the Swiss roll, we use the learned committor itself as the collective variable for adaptive sampling.  A similar approach was taken in \cite{rotskoff2022active} which generates new training data points for minimization of a cost function similar to \eqref{eq:0LagLoss} using 
% an umbrella sampling algorithm \cite{thiede2016eigenvector,dinner2020stratification}.
% % and replica exchange \cite{swendsen1986replica,fukunishi2002hamiltonian} for learning rare events recently \cite{rotskoff2022active}. 

% \section{Results}
% \subsection{Test System}
% insert background on SSW system.

\section{Predicting an atmospheric transition}\label{sec:HM}
As a demanding test of our method, we compute the committor and lead time for a model of sudden stratospheric warming (SSW), aiming to improve upon the benchmarks computed in \cite{finkel2021learning}. 
Like other models of geophysical flows, the dynamics are irreversible and the stationary distribution is unknown.  As a consequence, many competing approaches for computing the committor (e.g., \cite{li2019computing,khoo2019solving,li_semigroup_2022,roux_string_2021,roux2022transition}) are not applicable.
%SSW is an approximately biennial atmospheric event that disrupts the stratospheric polar vortex. 

Typical winter conditions in the Northern Hemisphere stratosphere support a strong polar vortex, fueled by a large equator-to-pole temperature gradient. Approximately once every two years, planetary waves rising from the troposphere impart a disturbance strong enough to weaken and destabilize the vortex, in some cases splitting it in half. Such events cause stratospheric temperatures to rise by about $50^\circ$C over several days, affecting surface weather conditions for up to three months. The polar vortex is dynamically coupled to the midlatitude (tropospheric) jetstream, which sometimes weakens in response to SSW. This can engulf the midlatitudes in Arctic air and alter storm tracks, bringing severe weather conditions to unprepared locations. Predicting SSW events is therefore a prime objective in subseasonal-to-seasonal weather prediction, but their abruptness poses a real challenge. For a review of SSW observations, predictability and modeling, see \cite{Baldwin2021sudden} and references therein.

We consider the Holton-Mass model \cite{Holton1976stratospheric}, augmented by time-dependent stochastic forcing as in \cite{finkel2021learning} to represent unresolved processes and excite transitions between the strong and weak vortices. Despite the simplicity of the model relative to state-of-the-art climate models, these transitions capture essential features of SSW such as the rapid upward burst of wave activity mediated by the ``preconditioned'' vertical structure of zonal-mean flow \cite{Sjoberg2014stratospheric,Maher2019model}.   We briefly describe the model here, but refer the interested reader to \cite{finkel2021learning,finkel2020path, Holton1976stratospheric,Christiansen2000} for additional background and details. 

The model domain is the region of the atmosphere north of 30$^\circ$ and above the altitude of $z\approx10$ km (the tropopause). The Holton-Mass model describes stratospheric flow in terms of a wave-mean flow interaction between two physical fields. The mean flow refers to the zonal-mean zonal wind $\overline{u}(y,z,t)$: the horizontal wind velocity component in the east-west (zonal) direction, averaged over a ring of constant latitude (zonal-mean, denoted by the overbar). The spatial coordinate $y$ denotes the north-south (meridional) distance from the latitude line $\phi_0=60^\circ$, i.e., $y=a(\phi-\phi_0)$, where $a$ is the Earth's radius and $\phi$ is the latitude. The wave refers to the perturbation streamfunction $\psi'(x,y,z,t)$: the deviation from zonal mean (denoted by a prime symbol) of the geostrophic streamfunction, which is proportional to the potential energy of a given air parcel. Holton and Mass worked with the following ansatz for the interaction:
\begin{equation}\label{eq:holton_mass_ansatz}
\begin{aligned}
    \overline{u}(y,z,t)&=U(z,t)\sin(\ell y)\\
    \psi'(x,y,z,t)&=\text{Re}\{\Psi(z,t)e^{ikx}\}e^{z/2H}\sin(\ell y)
\end{aligned}
\end{equation}
where $k=2/(a\cos\,60^\circ)$ and $\ell=3/a$ are zonal and meridional wavenumbers, and $H=7$ km is a scale height. The equations in \eqref{eq:holton_mass_ansatz} prescribe the horizontal structure entirely, so the model state space consists of $U(z,t)$ and $\Psi(z,t)$, the latter being complex-valued.  Insertion of \eqref{eq:holton_mass_ansatz} into the quasigeostrophic potential voriticity equation yields a system of two coupled PDEs.  Following \cite{finkel2021learning, Holton1976stratospheric,finkel2020path,Christiansen2000}, we discretize the PDEs along the $z$ dimension in 27 layers.  After enforcing boundary conditions, this results in a 75-dimensional state space:
\begin{equation}\label{eq:holton_mass_state}
\begin{aligned}
    X_t=[&\text{Re}\{\Psi(\Delta z,t)\},...,\text{Re}\{\Psi(25\Delta z,t)\},\\
   &\text{Im}\{\Psi(\Delta z,t)\},...,\text{Im}\{\Psi(25\Delta z,t)\}, \\
   &U(\Delta z,t),...,U(25\Delta z,t)].
\end{aligned}
\end{equation}

\begin{figure}[btp]
\begin{center}
\includegraphics[scale=0.5]{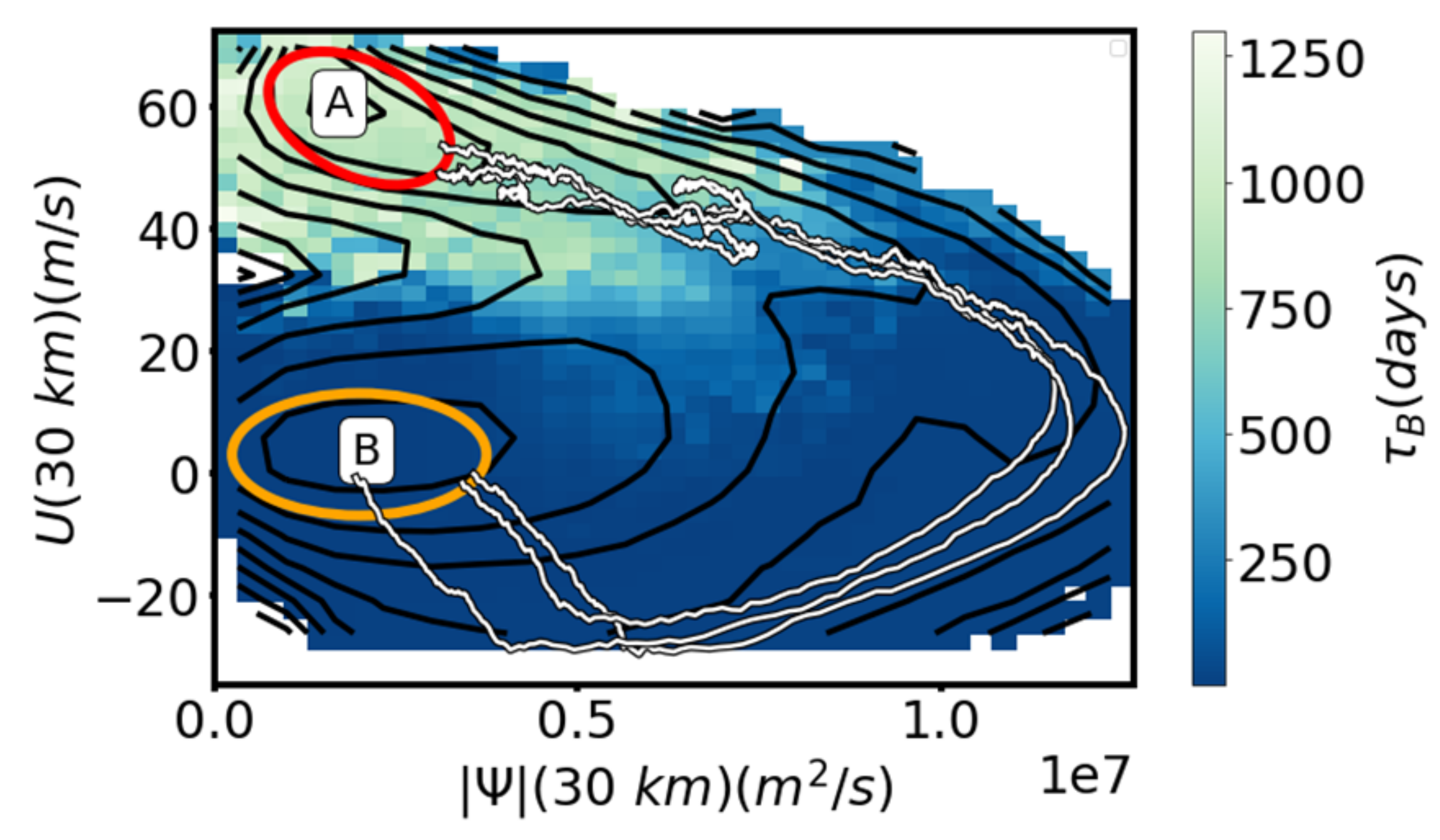}
\end{center}
\caption{\label{fig:HMIntro}
Illustration of some key properties of the Holton Mass model relevant to the prediction problems considered here.  Red and yellow ellipses approximately mark the projections of states $A$ and $B$, respectively, on the collective variables. The background color shows the average time to hit state $B$, clipped to a maximum of 1300 days to show detail.  Black contours show the negative logarithm of the stationary density marginalized on these collective variables.  Three transition paths harvested from a long simulation are shown in white.
}
\end{figure}

The two states of interest in this model are a strong polar vortex, with large positive $U(z,t)$ (meaning eastward wind, marked as state $A$ in Figure \ref{fig:HMIntro}), and a weak polar vortex, with a weak wind profile in which $U(z,t$) sometimes dips negative (marked as state $B$ in Figure \ref{fig:HMIntro}). Specifically, we define $A$ and $B$ as spheres centered on the model's two stable fixed points $(\Psi_{\mathbf{a}},U_{\mathbf{a}})$ and $(\Psi_{\mathbf{b}},U_{\mathbf{b}})$ in the 75-dimensional state space. 
The two spheres have radii of 8 and 20 respectively, with distances measured in the non-dimensionalized state space specified in \cite{finkel2021learning}.  In physical units, these correspond to the ellipsoids
\begin{align}
    A &= \bigg\{\Psi,U:\frac{\lVert\Psi-\Psi_{\mathbf{a}}\rVert^2}{(7.2\times10^5\text{ m}^2/\text{s})^2}+\frac{\lVert U-U_{\mathbf{a}}\rVert^2}{(2.9\text{ m/s})^2}\leq8^2\bigg\}\\
    B &= \bigg\{\Psi,U:\frac{\lVert\Psi-\Psi_{\mathbf{b}}\rVert^2}{(7.2\times10^5\text{ m}^2/\text{s})^2}+\frac{\lVert U-U_{\mathbf{b}}\rVert^2}{(2.9\text{ m/s})^2}\leq20^2\bigg\}
\end{align}
where $\lVert\cdot\rVert$ is the complex vector 2-norm.
%,  is the state vector of fixed point $\mathbf{a}$, and  is the state vector of fixed point $\mathbf{b}$. 

Figure \ref{fig:HMIntro} illustrates the key features of this model relevant to the prediction problems we consider here.  We see that the average time to reach $B$ starting from $A$ is over 1000 days, which is substantially longer than the longest lag times we consider here ($\leq$ 10 days).  We can also see that the transition paths do not proceed through the saddlepoint of the effective free energy (i.e., the negative logarithm of the stationary density, marked by the contours), indicating that dynamical, non-diffusive, irreversible dynamics are important. 
%SHORTER TRY1: 
{ Specifically, the transition path can be roughly divided into two stages: a ``preconditioning'' phase, in which the vortex gradually weakens, followed by an upward burst of wave activity that rips the vortex apart. Most of the committor's increase happens during the preconditioning phase, which siphons enstrophy (that is, squared vorticity, a measure of vortex strength that is conserved in the absence of dissipation) away from the mean flow and into the wave activity. The wave eventually dissipates, but only after its magnitude $|\Psi|$ bypasses the saddlepoint (Figure~\ref{fig:HMIntro}). 
See \cite{finkel2021exploring,Yoden1987_dyn} for further discussion.} 

%In addition, this system is highly non-reversible as the drift is not a gradient of a potential, and the exact stationary distribution is unknown.

\begin{figure}[btp]
\begin{center}
\includegraphics[scale=0.55]{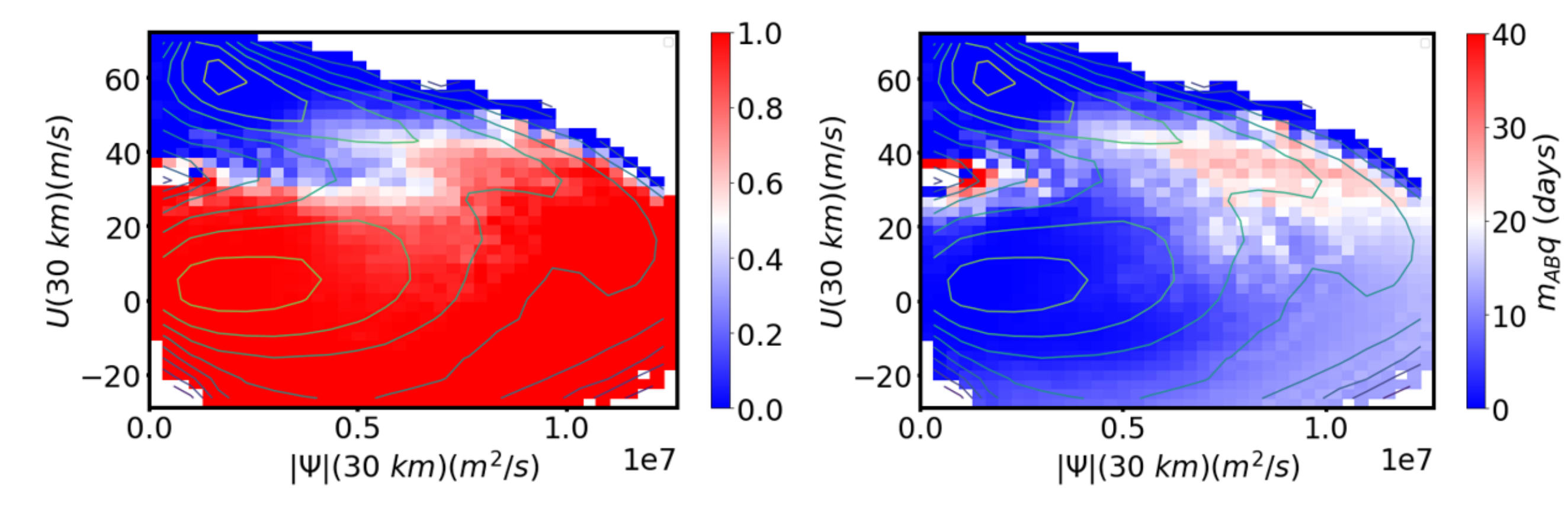}
\end{center}
\caption{\label{fig:Refs}
Reference statistics for the Holton-Mass model. (left) Committor and (right) lead time computed from a long trajectory and projected onto $U(30\ \text{km})$ and $|\Psi|(30\  \text{km})$.  Colors show reference statistics, and contours show the effective free energy.}
\end{figure}

To generate an initial data set, we sampled 30,000 points uniformly in $U(30\  \text{km})$ and $|\Psi|(30\ \text{km})$ from a long (50,000 days) trajectory and ran two ten-day trajectories from each starting point. Simulation details are reported in \cite{finkel2021learning}.  We simulated with a time step of 0.005 days, and saved the state of the system every 0.1 days.  To validate our neural network results, we use a long trajectory of 500,000 days to compute
\begin{equation}\label{eq:qhat}
    \langle q(s)\rangle=\mathbb{E}[\mathbbm{1}_B(X(\tau))|u_{\theta^*}(X(0))\in [s,s+\Delta s]]\ \text{for}\ 
s\in[0,1].
\end{equation}
where $\theta^*$ are the parameters obtained from solving \eqref{eqn:Min}. This is the mean reference committor over the isocommittor surfaces from the neural network function.  A perfect prediction corresponds to $\langle q(s)\rangle=s$.  We use a similar construction for $m_{AB}q$, which we denote $\langle[m_{AB}q](s)\rangle$.  For the committor, we take the overall error to be
\begin{equation}
    \text{Error}=\sqrt{\int_0^1(\langle q(s)\rangle-s)^2ds}
\end{equation}
Because the lead time does not have a fixed range { and scales exponentially with the noise}, for it, we instead compute the relative error
\begin{equation}
    \sqrt{\int_0^{40}(\langle[m_{AB}q](s)\rangle-s)^2/s^2ds}.
\end{equation}
Figure \ref{fig:Refs} shows the reference committor and lead time projected onto the collective variables.  

\begin{figure}[btp]
\begin{center}
\includegraphics[scale=0.5]{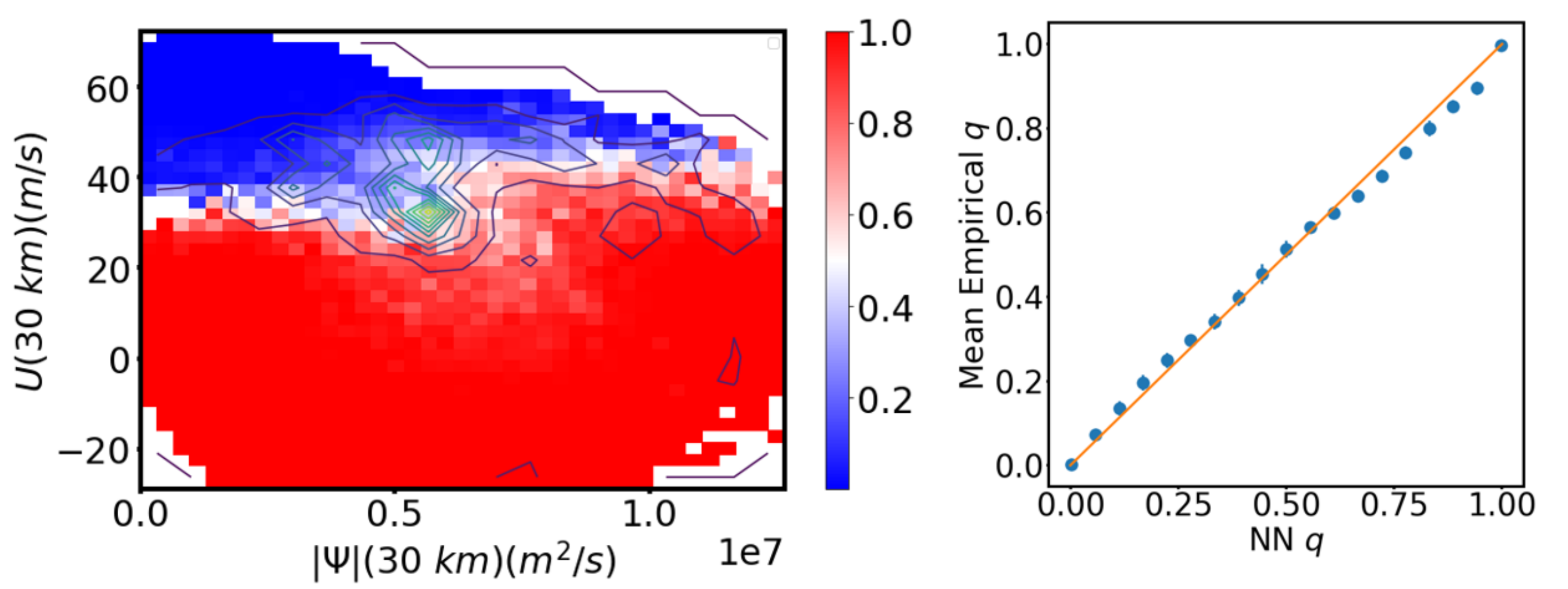}
\end{center}
\caption{\label{fig:QValidate}
Committor for the Holton-Mass model. (left) Colors show predictions, and contour lines show the density of added points from the adaptive sampling scheme described in Section \ref{sec:adapt}.  (right) Comparison of predicted and reference committors. Symbols show \eqref{eq:qhat}.  Error bars show the standard deviation from networks trained on ten separate data sets resulting from the adaptive sampling scheme.}
\end{figure}

%We tested our training procedure on both the forward committor and conditional mean first passage time problems. We use a fully-connected feed-forward architecture with ten layers containing 50 nodes each.

\begin{figure}[btp]
\begin{center}
\includegraphics[scale=0.5]{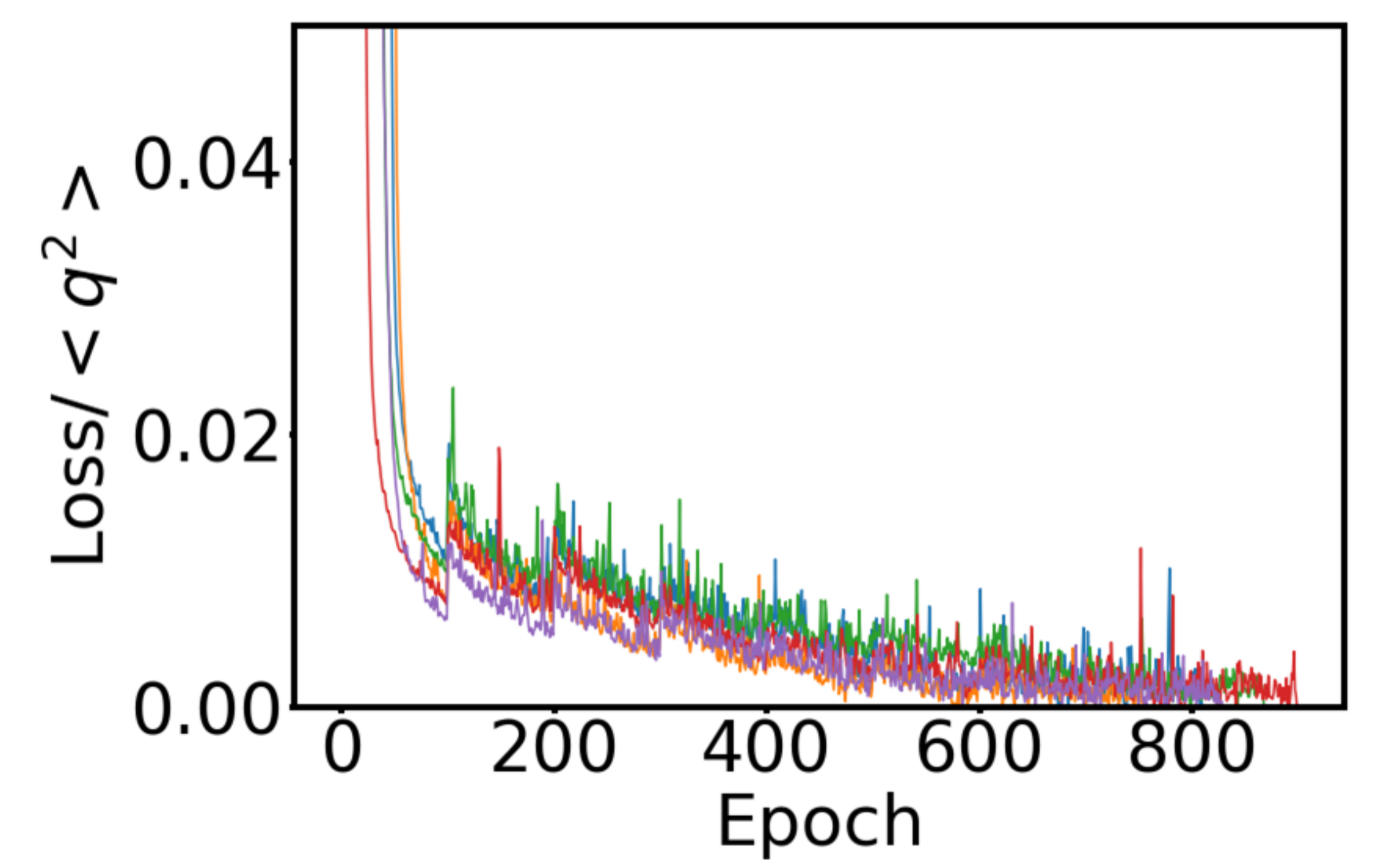}
\end{center}
\caption{\label{fig:Epochs}
The value of the loss as the training progresses for several replicates.   We add data adaptively every 100 epochs and halt training when the loss goes below zero. Synchronized spikes in the loss function result from adding data where the loss is high.}
\end{figure}

Figure \ref{fig:QValidate}(left) shows results for the committor obtained with the adaptive sampling method.  As collective variables in the adaptive sampling scheme we use  $\xi = (U(30\  \text{km}),|\Psi|(30\ \text{km}))$.  The space
 between $U(30\  \text{km}) = -29 \text{ and } 72.5\  \text{m}/\text{s}$ and between
$|\Psi|(30\ \text{km}) =0 \text{ and } 1.26\times 10^7\  \text{m}/\text{s}^2$
is partitioned into a $20\times 20$ grid of bins.
 We choose this collective variable space because it is physically intuitive, coming directly from the model's state space, and because it resolves SSW events well. 
%  This allows us to see the variability across paths.
 Physically, $U$ measures the strength of the vortex while $|\Psi|$ measures the strength of the disruptive wave. Their coupling is key to the nonlinear dynamics of the model.
We begin with 50,000 pairs of short trajectories and add 22,000 new pairs of ten-day trajectories every 100 epochs for a total of 10 cycles.  Thus the final number of trajectory pairs is 270,000.  We take $\lambda=10$ in \eqref{eq:approxcost}.  The network architecture is a fully  connected feed forward network with 75 inputs, 10 hidden layers of width 50, with ReLU activation functions, and an output layer with a single sigmoid activation function. We stop training between each addition of data whenever the loss goes below zero (Figure \ref{fig:Epochs}).   Networks for the lead time have the same structure, except that they have a quadratic output layer.  The contour lines in Figure \ref{fig:QValidate}(left) indicate the density produced at the end of the training by the adaptive sampling procedure.  The method concentrates new samples in the transition region without being given any information about its location.  The method  identifies the transition region on the fly.  

\begin{figure}[btp]
\begin{center}
\includegraphics[scale=0.5]{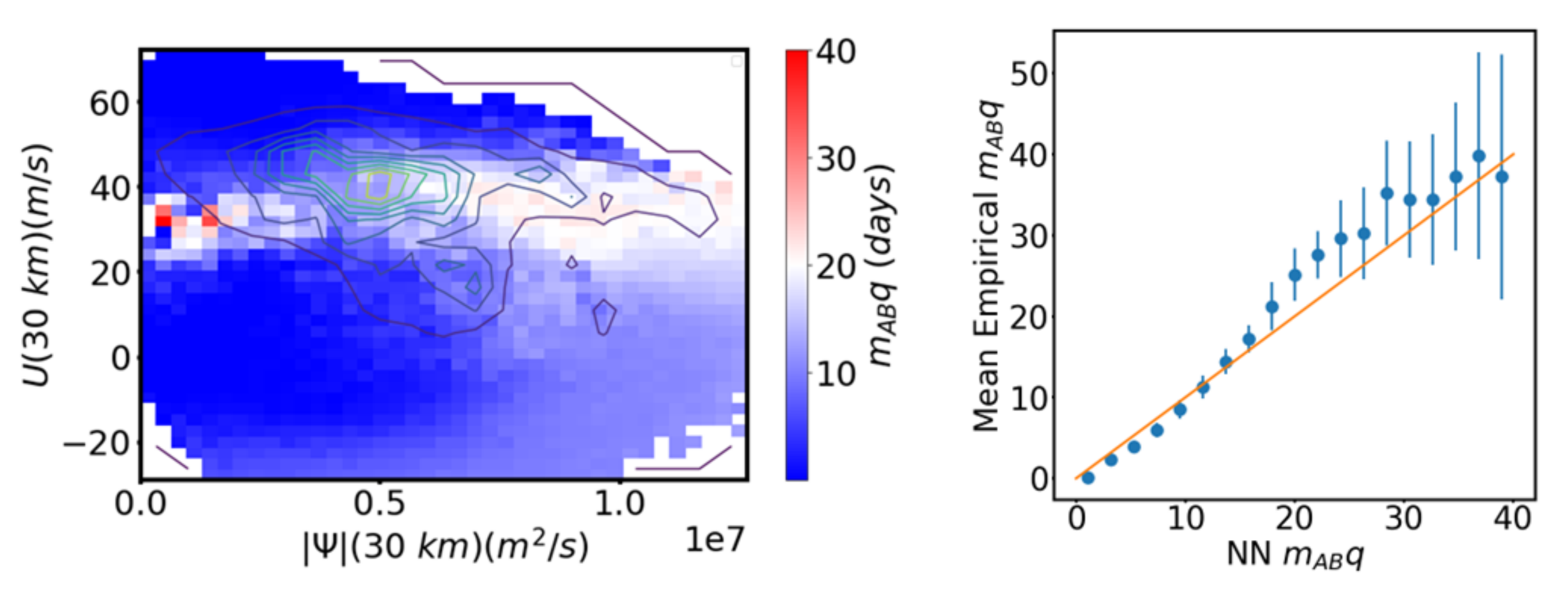}
\end{center}
\caption{\label{fig:CMFPTValidate}
Lead time for the Holton-Mass model.  (left) Colors show predictions, and contour lines show the density of points in the data set obtained from the adaptive sampling scheme for the committor.  (right)  Comparison of predicted and reference lead times. Symbols show $\langle[m_{AB}q](s)\rangle$.  Error bars show the standard deviation from networks trained on ten indpendent data sets resulting from the adaptive sampling scheme used to train the committor.}
\end{figure}

To validate the results, we trained ten networks on the data set produced by the adaptive sampling method and computed \eqref{eq:qhat} (Figure \ref{fig:QValidate}(right)).  The error bars show the standard deviation in $\langle q(s)\rangle$.  We see that the training is robust and consistently able to produce good estimates of the committor.  We used the data set obtained from the adaptive sampling scheme for the committor to train the neural network to predict the lead time (Figure \ref{fig:CMFPTValidate}).  Once again, we find that the method consistently produces good results compared with estimates from a long trajectory. { We expect the errors in Figure \ref{fig:CMFPTValidate} to be larger than those in Figure \ref{fig:QValidate} because the estimated committor is used in the loss function for the lead time (as discussed below \eqref{eqn:FK}), allowing errors to compound.}

\begin{figure}[btp]
\begin{center}
    \includegraphics[scale=0.5]{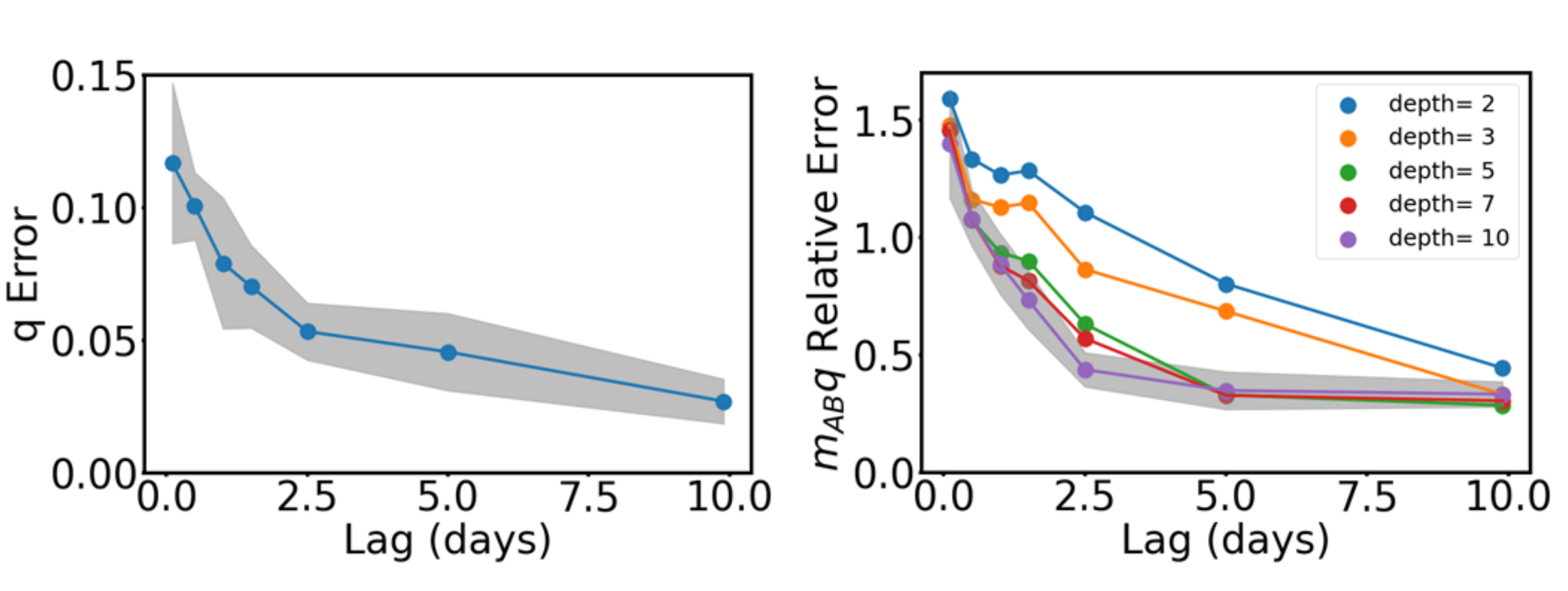}
\end{center}
\caption{\label{fig:LagError}
Dependence of performance as key hyperparameters are varied.  (left) RMS error of the committor as a function of lag time.  (right) Relative error in the product used to solve for the lead time as a function of the network depth and lag time. Shading shows the standard deviation from networks trained on ten indpendent data sets resulting from the adaptive sampling scheme used to train the committor; on the right, shading is only shown for the deepest network for clarity.}
\end{figure}

Finally, we determined how the performance of our method depended on key hyperparameters. To elucidate trends, we trained 10 networks on the data set produced by our adaptive sampling method.  Figure \ref{fig:LagError} shows the error in our scheme as the lag time is increased.  As we observed in the case of the rugged M\"uller-Brown potential, the error decreases as the lag time increases.  We note that as the lag time goes to infinity, all trajectories reach $A\cup B$, and the algorithm reduces to nonlinear regression of point estimates of the conditional expectation of being in state $B$ (see \eqref{eqn:qdef}).  We also investigated the dependence of the performance on the network depth, as shown in the right panel  of Figure \ref{fig:LagError}. We found that deeper networks were able to achieve low errors at intermediate lag times, although there was relatively little sensitivity to this hyperparameter at short and long lag times.  

\section{Conclusions}\label{sec:conc}
In this work, we have proposed a machine learning method for solving prediction problems given a data set of short trajectories.  By computing conditional expectations that solve Feynman-Kac equations rather than trying to learn the full dynamical law, we reduce the scope of the problem and hence render it more tractable.  
{
Our method has a number of advantages over existing ones:
\begin{itemize}
    \item it allows computation of any statistic that can be cast in Feynman-Kac form;
    \item it does not require explicit knowledge of either the model underlying the data or its dynamics  (e.g., the form of the generator and its parameters, such as the diffusion tensor);
    \item it allows for use of arbitrary lag times;
    \item it allows use of an arbitrary sampling distribution;
    \item it does not require microscopic reversibility.
\end{itemize}
}

We illustrate these advantages using two numerical examples.
Using a three-dimensional model for which we can compute an accurate reference solution, we show that our method using short trajectory data is often more robust than related methods that instead use the differential operator form of the Feynman-Kac equation \cite{li2019computing,khoo2019solving,li_semigroup_2022,rotskoff2022active}. 
% This improvement results from the stochastic sampling for a nonzero lag times averaging out the high frequency modes of the potential. 
With the same model, we demonstrate the importance of having data in the low probability regions between metastable states { and adequately weighting it against the data in the high probability regions.} We propose a simple adaptive sampling scheme that allows us to add data { so as to target the largest contributions to the loss} during training.  Finally, we show that we can compute key statistics for a 75-dimensional model of SSW events { (not just the committor but also the lead time)} from trajectories that are significantly shorter than the times between events.

Our method opens new possibilities for the study of rare events using experimental and observational data. For example, data sets of short trajectories generated by weather forecasting centers can be analysed by our method to study extreme weather and climate events~\cite{finkel2022revealing}. However, the requirement that two trajectories be generated from each initial condition poses an obstacle to application of our method to many other data sets. 
% This is not a severe cost in terms of serial computation time since the trajectories are independent and can be run in parallel. 
% However, it prohibits analysis of most experimental data sets. 
Future work will focus on relaxing this restriction.

%  Similarly, the ``analogue method'' of weather forecasting predicts the weather from a given initial condition by averaging over trajectories from similar initial conditions from the historical record (analogues). The analogue method is old \cite{Dool1989new}, but is enjoying a revival with the availability of larger data sets, especially for the prediction of extremes \cite{Chattopadhyay2020analog,Lucente2021coupling,Krouma2022assessment,Krouma2022ensemble}, 

\section*{Acknowledgments}

We wish to thank Adam Antoszewski, Michael Lindsey, Chatipat Lorpaiboon, and Robert Webber for useful discussions and the Research Computing Center at the University of Chicago for computational resources. This work was supported by National Institutes of Health award R35 GM136381 and National Science Foundation award DMS-2054306. J.F. was supported at the time of writing by the U.S. DOE, Office
of Science, Office of Advanced Scientific Computing Research, Department of Energy Computational
Science Graduate Fellowship under Award Number DE-SC0019323. % J.F. also acknowledges current support from the MIT Climate Grand Challenge on Weather and Climate Extremes

\bibliographystyle{unsrt}
\bibliography{arxiv_v2.bbl}

\end{document}